\documentclass[12pt,hidelinks]{article}
\pdfoutput=1
\usepackage{mathrsfs,amsmath,epsfig,amssymb,color,graphicx,bm}
\usepackage{natbib,subcaption}
\usepackage[colorlinks,linkcolor=black,citecolor=black,urlcolor=black]{hyperref}
\usepackage[margin=1in]{geometry}
\usepackage[symbol]{footmisc}
\usepackage[ruled,linesnumbered]{algorithm2e}
\usepackage[figuresright]{rotating}
\usepackage{multirow}
\graphicspath{{figures/}}
\allowdisplaybreaks

\renewcommand{\theequation}{\arabic{section}.\arabic{equation}}

\newtheorem{theorem}{Theorem}%[section]
\newtheorem{lemma}{Lemma}%[section]

\newcommand{\hbeta}{{\hat{ \boldsymbol \beta}}}
\newcommand{\bbeta}{{\boldsymbol \beta}}
\newcommand{\tbeta}{\tilde{\boldsymbol \beta}}

\newcommand{\cbeta}{\check{\boldsymbol \beta}}

\newcommand{\bSigma}{{\boldsymbol \Sigma}}

\begin{document}

\centerline {\Large\bf Renewable Learning for  Multiplicative Regression}
\centerline {\Large\bf       with Streaming Datasets }
\vspace*{0.2in}

\centerline{ {\large Tianzhen Wang$^{1}$, Haixiang Zhang$^{1*}$ and Liuquan Sun$^{2}$}}
 \vspace*{0.1in}

\centerline{\it  \small $^{1}$Center for Applied Mathematics, Tianjin University, Tianjin 300072, China}
\centerline{\it  \small $^{2}$Academy of Mathematics and Systems Science, Chinese Academy of Sciences, Beijing 100190, China}

\vspace{1cm}
\footnotetext[1]{Corresponding author: haixiang.zhang@tju.edu.cn (Haixiang Zhang)}

\begin{abstract}
When large amounts of data continuously arrive in streams, online updating is an effective way to reduce storage and computational burden. The key idea of online updating is that the previous estimators are sequentially updated only using the current data and some summary statistics of historical raw data. In this article, we develop a renewable learning method for a multiplicative regression model with streaming data, where the parameter estimator based on a least product relative error criterion is renewed without revisiting any historical raw data. Under some regularity conditions, we establish the consistency and asymptotic normality of the renewable estimator. Moreover, the theoretical results confirm that the proposed renewable estimator achieves the same asymptotic distribution as the least product relative error estimator with the entire dataset. Numerical studies and two real data examples are provided to evaluate the performance of our proposed method.


\end{abstract}

\section{Introduction}
With the rapid development of data collecting and storage technologies, the sizes of available datasets have grown rapidly during recent years. In the era of big data, it is common that  datasets  continuously arrive in streams or  large chunks. Faced with this kind of large-scale streaming dataset, many  conventional statistical methods are challenging mainly due to (i) the entire dataset is too large to be held in a general computer's memory; (ii) the historical data may no longer be accessible due to the storage burden or privacy limit. The online updating method is effective to address the two challenges, because it only needs the current block data and some summary statistics of previous data instead of historical raw data. To be more specific, the primary advantage of online updating method is that it does not require to access historical data, while it is able to provide real-time inference for making decisions. In the literature, many efforts have been devoted to develop online updating methods towards streaming datasets. For example,  \cite{schifano2016online} proposed a cumulative estimating equation (CEE) estimator and a cumulatively updated estimating equation (CUEE) estimator with streaming datasets. \cite{lee2020online} studied an online updating method to correct the bias due to covariate measurement error in the framework of linear models. \cite{luo2020renewable} developed an incremental updating algorithm to analyze streaming data for generalized linear model. \cite{lin2020unified} established a unified framework of renewable weighted sums for various online updating estimations with streaming datasets. \cite{xue2020online} proposed an online updating-based test to evaluate the proportional hazards assumption with streaming survival data. \cite{wu2021online} proposed an online updating method of survival analysis under the Cox proportional hazards model. \cite{luo2021multivariate} studied a multivariate online regression analysis with heterogeneous streaming data. \cite{lu2021online} studied a homogenization strategy for heterogeneous streaming data. \cite{hector2021parallel} proposed a new big data learning method by seamlessly integrating parallel data processing and online streaming paradigm. \cite{luo2021statistical} proposed an online debiased lasso method for high-dimensional generalized linear models with streaming data. \cite{shi2021online}  studied a novel framework for online causal learning. \cite{luo2022real} proposed an incremental learning algorithm to analyze streaming data with correlated outcomes based on quadratic inference function. \cite{wang2022renewable} proposed a novel online renewable strategy for quantile regression, among others.

In practice, we often meet with positive data in economic or biomedical studies. The multiplicative regression plays an important role in modeling this kind of positive data, such as stock prices or life times. In many applications, the relative error (e.g. stock price data), rather than error itself, is the major concern. The multiplicative regression is able to capture the size of  relative error. There have been several papers on the statistical analysis with multiplicative regression in the literature. e.g., \cite{chen2010least} proposed a least absolute relative error estimation criterion for multiplicative regression model. \cite{li2014empirical} considered an empirical likelihood approach towards constructing confidence intervals of the regression parameters in multiplicative regression model. \cite{chen2016least} proposed a least product relative error (LPRE) estimation criterion for multiplicative regression model. \cite{xia2016regularized} studied the variable selection for multiplicative regression model. Faced with large-scale streaming data with positive responses,  we propose a renewable learning method for multiplicative regression model. The main features of our approach are as follows: First, the renewable estimator and  its variance are sequentially updated only using the current data batch and some summary statistics of historical data, instead of the historical raw data. Therefore, the proposed method can deal with the computation and storage burden due to massive blocks of data. Second, the renewable estimator is statistically equivalent to the traditional LPRE estimator that based on the entire dataset, which implies that it achieves the same asymptotic distribution as the traditional LPRE estimator. Third, the computational speed of the proposed renewable learning method is much faster than the full data method.

The remainder of this article is organized as follows. In Section 2, we briefly review some notations for the multiplicative regression model with streaming data. In Section 3, we present a renewable estimation method and review two sequential updating methods. Section 4 investigates the theoretical properties of the proposed renewable estimator. In Section 5, we conduct some numerical simulations to evaluate the performance of our method. Section 6 presents two illustrative real data examples. In Section 7, we give some conclusions and future research topics.  All proofs are given in the Appendix.

\section{Model and Notations}
\newcounter{0}
We consider the following multiplicative regression model \citep{chen2010least},
\begin{align}\label{Model1}
Y_i=\exp(\bbeta^T\mathbf{X}_i)\epsilon_i,
\end{align}
where $Y_i$ is a positive response variable, $\mathbf{X}_i\in\mathbb{R}^p$ is a vector of covariates with the first component being 1 (intercept), $\bbeta=(\bbeta_1,\ldots,\bbeta_p)^T$ is a vector of regression parameters, and $\epsilon_i>0$ is an error term, $i=1\ldots N$. %The multiplicative regression model is widely used when analyzing data with positive responses,such as incomes, stock prices and survival times, etc.
To estimate the parameters in model (\ref{Model1}), \cite{chen2016least} proposed a LPRE criterion
\begin{align*}
\ell(\mathbf{Y};\mathbf{X},\bbeta)=\sum_{i=1}^N\left\{Y_i \exp(-\bbeta^T\mathbf{X}_i)+Y_i^{-1}\exp(\bbeta^T\mathbf{X}_i)-2\right\},
\end{align*}
which is an infinitely differentiable and strictly convex function, where $\mathbf{Y}=(Y_1,\ldots, Y_N)^T$ and $\mathbf{X}=(\mathbf{X}_1,\ldots,\mathbf{X}_N)^T$.  Accordingly, the score function is given by $\mathbf{S}(\mathbf{Y};\mathbf{X},\bbeta)= \nabla_\bbeta\ell(\mathbf{Y};\mathbf{X},\bbeta)$, where $\nabla_\bbeta$ stands for the derivative of $\ell(\mathbf{Y};\mathbf{X},\bbeta)$ with respect to $\bbeta$. Specifically, the score function has the following explicit expression:
\begin{align*}
\mathbf{S}(\mathbf{Y};\mathbf{X},\bbeta)=\sum_{i=1}^N\left\{Y_i^{-1}\exp(\bbeta^T\mathbf{X}_i)-Y_i \exp(-\bbeta^T\mathbf{X}_i)\right\}\mathbf{X}_i.
\end{align*}
Denote the minimizer of $\ell(\mathbf{Y};\mathbf{X},\bbeta)$ as $\hbeta_N$,  satisfying $\mathbf{S}(\mathbf{Y};\mathbf{X},\hbeta_N)=0$. Due to the convexity of $\ell(\mathbf{Y};\mathbf{X},\bbeta)$, the Newton-Raphson method is usually adopted to obtain the traditional LPRE estimator.

Note that streaming data with positive response is very common in many fields such as bioinformatics \citep{wei1992accelerated,jin2003rank} and economic analysis \citep{teekens1972some}. This brings new research opportunities, but also comes with challenges of storing and analyzing such streaming data. To be more specific,   the storage burden is heavy due to large blocks data. Moreover, it is often computationally infeasible to perform statistical analysis due to the relatively limited computing resources at hand. Meanwhile, the previous data may be not accessible due to privacy concern. Therefore, it is desirable to develop a renewable learning method for the multiplicative regression model that does not require storing any historical individual-level data in the streaming data environment. Assume that $\mathcal{D}_1,\ldots,\mathcal{D}_b,\ldots$ are independent and identically distributed streaming datasets, where $\mathcal{D}_b=\{(\mathbf{X}_{ib},Y_{ib})\}_{i=1}^{n_b}$ is the $b$th dataset. Let $\mathcal{D}_b^*=\{\mathcal{D}_1,\ldots,\mathcal{D}_b\}$ denotes the cumulative data up to batch $b$ with $N_b=\sum_{k=1}^b n_k$. As mentioned by \cite{luo2020renewable}, the key idea of renewable estimation method is that a previous estimator is sequentially updated only using the current data batch $\mathcal{D}_b$ and some summary statistics of historical data batches. To deal with large-scale streaming data with positive response, we will propose a renewable learning method for the multiplicative regression model in next section.

\section{Methods}
\setcounter{equation}{0}
\subsection{Renewable Estimation}

Let $\hbeta_b$ and $\hbeta_b^*$ be the traditional LPRE estimators obtained from a single batch $\mathcal{D}_b$ and the entire cumulative dataset $\mathcal{D}_b^*$, respectively. Denote $\tbeta_b$ as a renewable estimator obtained from the current data batch $\mathcal{D}_b$ and some summary statistics of historical data batches $\mathcal{D}_{b-1}^*$, where an initial estimator with the first data batch is $\tbeta_1=\hbeta_1=\hbeta_1^*$. For $b=2,3,\ldots$, a previous estimator $\tbeta_{b-1}$ is sequentially updated to $\tbeta_b$ using the current data batch $\mathcal{D}_b$ and a summary statistic of previous data batches $\mathcal{D}_{b-1}^*$. To illustrate the proposed method, we denote the score function on data batch $\mathcal{D}_b$ as follows:
\begin{align*}
\mathbf{S}_b(\mathcal{D}_b,\bbeta)=\sum_{i\in\mathcal{D}_b}
\left\{Y_{i}^{-1}\exp(\bbeta^T\mathbf{X}_{i})-Y_{i} \exp(-\bbeta^T\mathbf{X}_{i})\right\}\mathbf{X}_{i},
\end{align*}
and its negative gradient matrix is
\begin{align*}
\mathbf{Q}_b(\mathcal{D}_b,\bbeta)=-\sum_{i\in\mathcal{D}_b}\left\{Y_{i} \exp(-\bbeta^T\mathbf{X}_{i})+Y_{i}^{-1}\exp(\bbeta^T\mathbf{X}_{i})\right\}\mathbf{X}_{i}\mathbf{X}_{i}^T.
\end{align*}

For simplicity, we first consider two data batches $\mathcal{D}_1$ and $\mathcal{D}_2$. For the first data batch $\mathcal{D}_1$, a LPRE $\hbeta_1$ is obtained by solving $\mathbf{S}_1(\mathcal{D}_1,\hbeta_1)=0$. When the second data batch $\mathcal{D}_2$ arrives, the traditional LPRE estimator $\hat{\bbeta}_2^*$ satisfies the following aggregated score equation,
\begin{align} \label{Eq3.1}
\mathbf{S}_1(\mathcal{D}_1,\hat{\bbeta}_2^*)+\mathbf{S}_2(\mathcal{D}_2,\hat{\bbeta}_2^*)=0.
\end{align}
However, solving equation (\ref{Eq3.1}) requires revisiting the previous data batch $\mathcal{D}_1$. To derive a renewable estimator that does not need to revisit $\mathcal{D}_1$, we take the first-order Taylor expansion of $\mathbf{S}_1(\mathcal{D}_1,\hat{\bbeta}_2^*)$ at the estimator $\tbeta_1$,
\begin{align*}
\mathbf{S}_1(\mathcal{D}_1,\tbeta_1)+\mathbf{Q}_1(\mathcal{D}_1,\tbeta_1)
(\tbeta_1-\hbeta_2^*)+O_p\left({\|\hbeta_2^*-\tbeta_1\|}^2\right)+\mathbf{S}_2(\mathcal{D}_2,\hat{\bbeta}_2^*)=0.
\end{align*}
If min$\{n_1,n_2\}$ is large enough, both $\hbeta_2^*$ and $\tbeta_1$ are consistent estimators of the true value $\bbeta_t$ \citep{chen2016least}. After ignoring the error term $O_p\left({\|\hbeta_2^*-\tbeta_1\|}^2\right)$, we can derive a renewable estimator $\tbeta_2$ satisfying

\begin{align*}
\mathbf{S}_1(\mathcal{D}_1,\tbeta_1)+\mathbf{Q}_1(\mathcal{D}_1,\tbeta_1)
(\tbeta_1-\tbeta_2)+\mathbf{S}_2(\mathcal{D}_2,\tbeta_2)=0.
\end{align*}
Due to the fact that $\mathbf{S}_1(\mathcal{D}_1,\tbeta_1)=0$, the renewable estimator $\tbeta_2$ satisfies the following estimating equation:
\begin{align}\label{Eq3.2}
\mathbf{Q}_1(\mathcal{D}_1,\tbeta_1)
(\tbeta_1-\tbeta_2)+\mathbf{S}_2(\mathcal{D}_2,\tbeta_2)=0.
\end{align}

In a similar way, the traditional LPRE estimator $\hat{\bbeta}_3^*$ satisfies the following aggregated score equation after data batch $\mathcal{D}_3$ arrives,
\begin{align*}
\mathbf{S}_1(\mathcal{D}_1,\hat{\bbeta}_3^*)+\mathbf{S}_2(\mathcal{D}_2,\hat{\bbeta}_3^*)+\mathbf{S}_3(\mathcal{D}_3,\hat{\bbeta}_3^*)=0.
\end{align*}
Taking the first-order Taylor expansion of $\mathbf{S}_1(\mathcal{D}_1,\hat{\bbeta}_3^*)$ and $\mathbf{S}_2(\mathcal{D}_2,\hat{\bbeta}_3^*)$ at $\tbeta_1$ and $\tbeta_2$, respectively, we obtain

\begin{align}\label{Eq3.3}
&\mathbf{S}_1(\mathcal{D}_1,\tbeta_1)+\mathbf{Q}_1(\mathcal{D}_1,\tbeta_1)(\tbeta_1-\hbeta_3^*)+O_p\left({\|\hbeta_3^*-\tbeta_1\|}^2\right)+\mathbf{S}_2(\mathcal{D}_2,\tbeta_2)\nonumber\\
&+\mathbf{Q}_2(\mathcal{D}_2,\tbeta_2)(\tbeta_2-\hbeta_3^*)+O_p\left({\|\hbeta_3^*-\tbeta_2\|}^2\right)+\mathbf{S}_3(\mathcal{D}_3,\hat{\bbeta}_3^*)=0.
\end{align}
The error terms $O_p\left({\|\hbeta_3^*-\tbeta_1\|}^2\right)$ and $O_p\left({\|\hbeta_3^*-\tbeta_2\|}^2\right)$ in (\ref{Eq3.3}) could be asymptotically ignored if min$\{n_1,n_2,n_3\}$ is large enough. Removing such error terms, it is straightforward to deduce that
\begin{align*}
\mathbf{S}_1(\mathcal{D}_1,\tbeta_1)+\mathbf{Q}_1(\mathcal{D}_1,\tbeta_1)(\tbeta_1-\hbeta_3^*)+\mathbf{S}_2(\mathcal{D}_2,\tbeta_2)
+\mathbf{Q}_2(\mathcal{D}_2,\tbeta_2)(\tbeta_2-\hbeta_3^*)+\mathbf{S}_3(\mathcal{D}_3,\hat{\bbeta}_3^*)=0.
\end{align*}
In view of $\mathbf{S}_1(\mathcal{D}_1,\tbeta_1)=0$ and (\ref{Eq3.2}), we have
\begin{align*}
\mathbf{Q}_1(\mathcal{D}_1,\tbeta_1)(\tbeta_1-\hbeta_3^*)-\mathbf{Q}_1(\mathcal{D}_1,\tbeta_1)
(\tbeta_1-\tbeta_2)+\mathbf{Q}_2(\mathcal{D}_2,\tbeta_2)(\tbeta_2-\hbeta_3^*)+\mathbf{S}_3(\mathcal{D}_3,\hat{\bbeta}_3^*)=0.
\end{align*}
By merging the terms $\mathbf{Q}_1(\mathcal{D}_1,\tbeta_1)(\tbeta_1-\hbeta_3^*)$ and $-\mathbf{Q}_1(\mathcal{D}_1,\tbeta_1)
(\tbeta_1-\tbeta_2)$, we obtain this expression:

\begin{align*}
\mathbf{Q}_1(\mathcal{D}_1,\tbeta_1)(\tbeta_2-\hbeta_3^*)+\mathbf{Q}_2(\mathcal{D}_2,\tbeta_2)(\tbeta_2-\hbeta_3^*)+\mathbf{S}_3(\mathcal{D}_3,\hat{\bbeta}_3^*)=0.
\end{align*}
Therefore, the renewable estimator $\tbeta_3$ is a solution to the following equation:

\begin{align*}
\mathbf{Q}_1(\mathcal{D}_1,\tbeta_1)(\tbeta_2-\tbeta_3)+\mathbf{Q}_2(\mathcal{D}_2,\tbeta_2)(\tbeta_2-\tbeta_3)+\mathbf{S}_3(\mathcal{D}_3,\tbeta_3)=0.
\end{align*}

Similarly, we introduce a renewable estimator $\tbeta_b$ satisfying the incremental estimating equation:

\begin{align}\label{Eq3.4}
\sum_{k=1}^{b-1}\mathbf{Q}_k(\mathcal{D}_k,\tbeta_k)
(\tbeta_{b-1}-\tbeta_b)+\mathbf{S}_b(\mathcal{D}_b,\tbeta_b)=0.
\end{align}
For  convenience, we denote the aggregated negative gradient matrix $\sum_{k=1}^{b}\mathbf{Q}_k(\mathcal{D}_k,\tbeta_k)$ as $\tilde{\mathbf{Q}}_b$. Based on (\ref{Eq3.4}), the renewable estimator $\tbeta_b$ can be easily solved via the following Newton-Raphson iterations,
\begin{align*}
\tbeta_b^{(m+1)}=\tbeta_b^{(m)}-\left\{\tilde{\mathbf{Q}}_{b-1}+\mathbf{Q}_b(\mathcal{D}_b,\tbeta_b^{(m)})\right\}^{-1}
\tilde{\mathbf{S}}_b^{(m)},
\end{align*}
where the  adjusted score $\tilde{\mathbf{S}}_b^{(m)}=\tilde{\mathbf{Q}}_{b-1}
(\tbeta_{b-1}-\tbeta_b^{(m)})+\mathbf{S}_b(\mathcal{D}_b,\tbeta_b^{(m)})$ is updated over iterations. To speed up the calculations, we may avoid
updating the negative gradient matrix $\mathbf{Q}_b(\mathcal{D}_b,\tbeta_b^{(m)})$ at each iteration. As suggested by \cite{luo2020renewable},  we obtain the following iterative formula by replacing $\tbeta_b^{(m)}$ with $\tbeta_{b-1}$ in $\mathbf{Q}_b(\mathcal{D}_b,\tbeta_b^{(m)})$,
\begin{align*}
\tbeta_b^{(m+1)}=\tbeta_b^{(m)}-\left\{\tilde{\mathbf{Q}}_{b-1}+\mathbf{Q}_b(\mathcal{D}_b,\tbeta_{b-1})\right\}^{-1}
\tilde{\mathbf{S}}_b^{(m)}.
\end{align*}

\subsection{Sequential Updating Methods}
Note that \cite{schifano2016online} proposed a general cumulative estimating equation (CEE) approach, which can be directly applied to the multiplicative regression model with streaming data. For comparison, we provide some details on the CEE estimator $\tbeta_b^{cee}$ for model (\ref{Model1}), which is given by

\begin{align*}
\tbeta_b^{cee}={(\tilde{\mathbf{Q}}_{b-1}^{cee}+\mathbf{Q}_b^{cee})}^{-1}(\tilde{\mathbf{Q}}_{b-1}^{cee}\tbeta_{b-1}^{cee}+
\mathbf{Q}_b^{cee}\hbeta_b),~~~ \tilde{\mathbf{Q}}_{b}^{cee}=\sum_{k=1}^b\mathbf{Q}_k^{cee},~~b=1,2,\ldots.
\end{align*}
Here $\tbeta_0^{cee}=\mathbf{0}_{p\times1}$, $\tilde{\mathbf{Q}}_{0}^{cee}=\mathbf{0}_{p\times p}$ and $\mathbf{Q}_b^{cee}=-\nabla_{\bbeta}\mathbf{S}_b(\mathcal{D}_b,\hbeta_b)$ is the negative gradient matrix of data batch $\mathcal{D}_b$.
For symbol simplicity, we denote
\begin{align*}
\mathbf{C}_b(\mathcal{D}_b,\bbeta)=\mathbf{S}_b(\mathcal{D}_b,\bbeta)\mathbf{S}_b(\mathcal{D}_b,\bbeta)^T=
\sum_{i\in\mathcal{D}_b}\left\{Y_{i}^{-1}\exp(\bbeta^T\mathbf{X}_{i})-Y_{i}\exp(-\bbeta^T\mathbf{X}_{i})\right\}^2\mathbf{X}_{i}\mathbf{X}_{i}^T,
\end{align*}
and $\mathbf{C}_b^{cee}=\mathbf{C}_b(\mathcal{D}_b,\hbeta_b)$. Set $\tilde{\mathbf{V}}_{0}^{cee}=\mathbf{0}_{p\times p}$,  from equation (18) of \citet{schifano2016online}, the variance of $\tbeta_b^{cee}$ is

\begin{align*}
&\tilde{\mathbf{V}}_{b}^{cee}=(\tilde{\mathbf{Q}}_{b-1}^{cee}+\mathbf{Q}_b^{cee})^{-1}
\left\{\tilde{\mathbf{Q}}_{b-1}^{cee}\tilde{\mathbf{V}}_{b-1}^{cee}(\tilde{\mathbf{Q}}_{b-1}^{cee})^T+\mathbf{Q}_b^{cee}\hat{\mathbf{V}}_b^{cee}
(\mathbf{Q}_b^{cee})^T\right\}\nonumber\\
&~~~~~~~~~~~~~~~~~~~~~~~\times\left\{(\tilde{\mathbf{Q}}_{b-1}^{cee}+\mathbf{Q}_b^{cee})^{-1}\right\}^T, ~~b=1,2,\ldots,
\end{align*}
where $\hat{\mathbf{V}}_b^{cee}=\left\{\mathbf{Q}_b^{cee}\mathbf{C}_b^{cee^{-1}}\mathbf{Q}_b^{cee^T}\right\}^{-1}$ presented by \cite{chen2016least} is the estimated variance of $\hbeta_b$ from the $b$th data batch.

To further reduce bias of the CEE estimator, a cumulatively updated estimating equation (CUEE) estimator was proposed by \cite{schifano2016online}.
%\begin{align}
%\cbeta_b={(\tilde{\mathbf{Q}}_{b-1}^{cee}+\mathbf{Q}_b^{cee})}^{-1}\left(\sum_{k=1}^{b-1}\mathbf{Q}_k^{cee}\cbeta_k+\mathbf{Q}_b^{cee}\hbeta_b\right)
%\end{align}
Denote $\mathbf{Q}_b^{cuee}=-\nabla_\bbeta\mathbf{S}_b(\mathcal{D}_b,\cbeta_b)$ and $\mathbf{C}_b^{cuee}=\mathbf{C}_b(\mathcal{D}_b,\cbeta_b)$, where $\cbeta_b$ is a CEE estimator. From equations (22) and (23) of \citet{schifano2016online}, with initial $\tilde{\mathbf{Q}}_{0}^{cuee}=\mathbf{Q}_{0}^{cuee}=\mathbf{0}_{p\times p}$ and $\tilde{\mathbf{V}}_{0}^{cuee}=\mathbf{0}_{p\times p}$, the CUEE estimator and its corresponding variance matrix  are
\begin{align*}
&\tbeta_b^{cuee}={(\tilde{\mathbf{Q}}_{b-1}^{cuee}+\mathbf{Q}_b^{cuee})}^{-1}\left\{\sum_{k=1}^{b-1}\mathbf{Q}_k^{cuee}\cbeta_k+
\mathbf{Q}_b^{cuee}\cbeta_b-\sum_{k=1}^{b-1}\mathbf{S}_k(\mathcal{D}_k,\cbeta_k)-\mathbf{S}_b(\mathcal{D}_b,\cbeta_b)\right\},
\end{align*}
and
\begin{align*}
&\tilde{\mathbf{V}}_{b}^{cuee}=(\tilde{\mathbf{Q}}_{b-1}^{cuee}+\mathbf{Q}_b^{cuee})^{-1}
\left\{\tilde{\mathbf{Q}}_{b-1}^{cuee}\tilde{\mathbf{V}}_{b-1}^{cuee}(\tilde{\mathbf{Q}}_{b-1}^{cuee})^T+\mathbf{Q}_b^{cuee}\hat{\mathbf{V}}_b^{cuee}
(\mathbf{Q}_b^{cuee})^T\right\}\nonumber\\
&~~~~~~~~~~~~~~~~~~~~~~~\times\left\{(\tilde{\mathbf{Q}}_{b-1}^{cuee}+\mathbf{Q}_b^{cuee})^{-1}\right\}^T,~~for ~~b=1,2,\ldots,
\end{align*}
where $\tilde{\mathbf{Q}}_{b}^{cuee}=\sum_{k=1}^b\mathbf{Q}_k^{cuee}$ and $\hat{\mathbf{V}}_b^{cuee}=\left\{\mathbf{Q}_b^{cuee}\mathbf{C}_b^{cuee^{-1}}\mathbf{Q}_b^{cuee^T}\right\}^{-1}$.

By \cite{schifano2016online}, the consistency of CEE and CUEE estimators are established under a strong regularity condition, i.e., the number of data batches $b$ is of order $O(n_k^j)$, for $j<1/3$ and all $k=1,\ldots,b$. However, this condition is not always valid for streaming data, because $n_k$ is typically small, but $b$ grows at a high rate. We will compare the proposed renewable estimator with the CEE and CUEE estimators via numerical simulations.

\section{Theoretical Properties}
\setcounter{equation}{0}

For notational simplicity, we denote
\begin{align}\label{Eq4.1}
\mathbb{C}(\bbeta)=\mathbb{E}_{\bbeta}\{\mathbf{S}(\mathbf{Y};\mathbf{X},\bbeta)\mathbf{S}(\mathbf{Y};\mathbf{X},\bbeta)^T\},
\end{align}
and
\begin{align}\label{Eq4.2}
\mathbb{Q}(\bbeta)=-\mathbb{E}_{\bbeta}\{\nabla_\bbeta\mathbf{S}(\mathbf{Y};\mathbf{X},\bbeta)\}.
\end{align}

To establish the consistency and asymptotic normality of the renewable estimator, we need the following regularity conditions.

\noindent{\bf (C.1)}  The true parameter $\bbeta_t$ lies in the interior of a compact set $\Theta\subset\mathbb{R}^p$.

\noindent{\bf (C.2)} The terms $\mathbb{C}(\bbeta)$ and $\mathbb{Q}(\bbeta)$ are positive-definite for all $\bbeta\in\mathbb{N}_\delta(\bbeta_t)$, where $\mathbb{N}_\delta(\bbeta_t)=\{\bbeta:\|\bbeta-\bbeta_t\|\leq\delta\}$ is a neighborhood around true value $\bbeta_t$, and $\delta$ is a positive constant.

\noindent{\bf  (C.3)}  $\sup_{\beta\in\Theta}\frac{1}{N}\sum_{i=1}^N\left\{Y_i\exp(-\bbeta^T\mathbf{X}_i) +Y_i^{-1}\exp(\bbeta^T\mathbf{X}_i)\right\}^2\|\mathbf{X}_i\|^2 = O_P(1)$.

%\noindent{\bf (C.4)} The $\mathbf{Q}_b(\mathcal{D}_b,\bbeta)$ is Lipschitz continuous with respect to $\bbeta$ in parameter space $\Theta$.

%\noindent{\bf (C.5)} $\mathbb{E}_\bbeta\{\|\mathbf{S}(Y;\mathbf{X},\bbeta)\|^2\}<\infty$ for all $\bbeta\in\Theta$.

Conditions (C.1) and (C.2) are regularity conditions, \citep[e.g.,][]{chen2016least}. Condition (C.3) is used to establish the consistency of the renewable estimator $\tbeta_b$, together with its asymptotic distribution. %Condition (C.4) implies that $\|\mathbf{Q}_b(\mathcal{D}_b,\bbeta_1)-\mathbf{Q}_b(\mathcal{D}_b,\bbeta_2)\|_2\leq M(\mathcal{D}_b)\|\bbeta_1-\bbeta_2\|_2$ for $\bbeta_1,\bbeta_2\in\Theta$, where $M(\mathcal{D}_b)$ is a vector of covariates $\mathbf{X}$ which are all bounded under a fixed design considered in this paper.

We first establish the consistency of the renewable estimator $\tbeta_b$ towards the true value $\bbeta_t$.

\begin{theorem}\label{th1}
If the conditions (C.1)-(C.3) hold and $N_b=\sum_{k=1}^b n_k\rightarrow\infty$, then the renewable estimator $\tbeta_b$ given in equation (\ref{Eq3.4}) is consistent to $\bbeta_t$, i.e. $\tbeta_b-\bbeta_t\stackrel{P}{\longrightarrow}0$.\textsf{}
\end{theorem}

To conduct statistical inference, we present the asymptotic normality of the renewable estimator $\tbeta_b$ in the following theorem.
\begin{theorem}\label{th2}
If the conditions (C.1)-(C.3) hold and $N_b=\sum_{k=1}^b n_k\rightarrow\infty$, then the renewable estimator $\tbeta_b$ has a mean-zero asymptotic normal distribution:
\begin{align*}
\sqrt{N_b}(\tbeta_b-\bbeta_t)\stackrel{{d}}{\longrightarrow} N(\mathbf{0},\mathbb{G}^{-1}(\bbeta_t)),
\end{align*}
where $ \stackrel{{d}}{\longrightarrow} $ denotes convergence in distribution, $\mathbb{G}(\bbeta_t)=
\mathbb{Q}^T(\bbeta_t)\mathbb{C}^{-1}(\bbeta_t)\mathbb{Q}(\bbeta_t)$, and $\mathbb{C}(\bbeta_t)$ and $\mathbb{Q}(\bbeta_t)$ are given in (\ref{Eq4.1}) and (\ref{Eq4.2}), respectively.
\end{theorem} \vspace{0.7cm}

It is worth mentioning that the asymptotic covariance of the renewable estimator $\tbeta_b$ given in Theorem~\ref{th2} is exactly the same as that of the traditional LPRE $\hbeta_b^*$ \citep[Theorem 3 in][]{chen2016least} based on the full data. This implies that the renewable estimator achieves the same efficiency as the traditional LPRE $\hbeta_b^*$. Using the aggregated  matrix $\tilde{\mathbf{Q}}_b=\sum_{k=1}^{b}\mathbf{Q}_k(\mathcal{D}_k,\tbeta_k)$ and $\tilde{\mathbf{C}}_b=\sum_{k=1}^{b}\mathbf{C}_k(\mathcal{D}_k,\tbeta_k)$, we calculate the estimated asymptotic covariance matrix as
\begin{align*}
\tilde{\bSigma_b}(\bbeta_t)=\left(N_b^{-1}\tilde{{\mathbf{G}}_b}\right)^{-1}= N_b\left(\tilde{\mathbf{Q}}_b^T \tilde{\mathbf{C}}_b^{-1}\tilde{\mathbf{Q}}_b\right)^{-1},
\end{align*}
where $\tilde{{\mathbf{G}}_b}=\tilde{\mathbf{Q}}_b^T \tilde{\mathbf{C}}_b^{-1}\tilde{\mathbf{Q}}_b$. The estimated asymptotic variance matrix for the renewable estimator $\tbeta_b$ is
\begin{align*}
 \tilde{\mathbf{V}}(\tbeta_b)=\frac{1}{N_b}\tilde{\bSigma_b}(\bbeta_t)=\left(\tilde{\mathbf{Q}}_b^T \tilde{\mathbf{C}}_b^{-1}\tilde{\mathbf{Q}}_b\right)^{-1}.
\end{align*}

In Algorithm~\ref{alg1}, we summarize the procedure of the renewable estimation for multiplicative regression with streaming data.

\begin{algorithm}
    \caption{Renewable Estimation}
    \label{alg1}
    \KwIn{Sequentially arrived datasets $\mathcal{D}_1,\ldots,\mathcal{D}_b,\ldots$ \;}
    \KwOut{$\tbeta_b$ and $\tilde{\mathbf{V}}(\tbeta_b)$, for $b=1,2,\ldots$ \;}
    Initialize: set initial values $\tbeta_{init}=\mathbf{0}$, $\tilde{\mathbf{Q}}_0=\mathbf{0}_{p\times p}$ and $\tilde{\mathbf{C}}_0=\mathbf{0}_{p\times p}$\;
    \For{$b=1,2,\ldots$}
    {
      Load the dataset $\mathcal{D}_b$\;
      \Repeat{convergence}
      {
        $\tbeta_b^{(m+1)}=\tbeta_b^{(m)}-\left\{\tilde{\mathbf{Q}}_{b-1}+\mathbf{Q}_b(\mathcal{D}_b,\tbeta_{b-1})\right\}^{-1}
        \left\{\tilde{\mathbf{Q}}_{b-1}(\tbeta_{b-1}-\tbeta_b^{(m)})+\mathbf{S}_b(\mathcal{D}_b,\tbeta_b^{(m)})\right\}$\;

      }
      Update $\tilde{\mathbf{Q}}_b=\tilde{\mathbf{Q}}_{b-1}+\mathbf{Q}_b(\mathcal{D}_b,\tbeta_b)$ and $\tilde{\mathbf{C}}_b=\tilde{\mathbf{C}}_{b-1}+\mathbf{C}_b(\mathcal{D}_b,\tbeta_b)$\;
      Calculate $\tilde{\mathbf{V}}(\tbeta_b)=\left\{\tilde{\mathbf{Q}}_b^T \tilde{\mathbf{C}}_b^{-1}\tilde{\mathbf{Q}}_b\right\}^{-1} $\;
      Release the dataset $\mathcal{D}_b$ from the memory.
    }
    Output $\tbeta_b$ and $\tilde{\mathbf{V}}(\tbeta_b)$, for $b=1,2,\ldots$.
    %\end{algorithmic}
\end{algorithm}

\section{Simulation Study}

In this section, we conduct some simulations to demonstrate the effectiveness of our proposed method. The true value of $\bbeta$ is chosen as $\bbeta_t= (0.2,-0.2,0.2,-0.2,0.2)^T$.  Denote the covariate $\mathbf{X} = (1, \tilde{\mathbf{X}}^T)^T$ with $\tilde{\mathbf{X}} = (X_1,\ldots,X_4)^T$, i.e. $p=5$. We consider two cases for the error term: $\log(\epsilon)$ follows $N(0, 1)$, and $\log(\epsilon)$ follows uniform distribution over $(-2,2)$. Moreover, we choose four cases for the covariate $\tilde{\mathbf{X}}$,

\begin{description}
 \item[Case 1.] $\tilde{\mathbf{X}} \sim N(\mathbf{0},\bSigma)$, where $\bSigma_{ij}= 0.5^{|i-j|}$.
 \item[Case 2.] $\tilde{\mathbf{X}} \sim 0.5N(\mathbf{1},\bSigma)+0.5N(\mathbf{-1},\bSigma)$, where $\bSigma_{ij}= 0.5^{|i-j|}$.
 \item[Case 3.] $\tilde{\mathbf{X}} \sim t_5(\mathbf{0},\bSigma)$, i.e., $\tilde{\mathbf{X}}$ follows a multivariate $t$ distribution with degree of freedom $\chi=5$ and  covariance matrix $\bSigma_{ij}= 0.5^{|i-j|}$.
 \item[Case 4.] $\tilde{\mathbf{X}} = (X_1,\ldots,X_4)^T$, where $X_i$'s are independently and  identically  distributed  exponential random variables with the probability density function $f(x)=e^{-x}$.
\end{description}

For comparison, we compare our renewable estimator with three competing estimators, which include

\begin{description}
 \item[(i)] the traditional LPRE estimator obtained from the entire data \citep{chen2016least};
 \item[(ii)] the CEE estimator \citep{schifano2016online};
 \item[(iii)] the CUEE estimator \citep{schifano2016online}.
\end{description}

These methods are compared from two aspects of estimation efficiency and computation speed. The results for
estimation efficiency include : the estimated bias (BIAS) given by the sample mean of the estimates minus the true
value $\bbeta_t$, the sampling standard error (SSE) of the estimates, the mean of the estimated standard errors (ESE) and the empirical 95\% coverage probabilities (CP) towards the true value $\bbeta_t$. In addition, the computation efficiency is assessed by computation time (C.Time) and running
time (R.Time), where C.Time includes time on data loading and algorithm execution, and R.time only accounts time for algorithm execution. All the simulation results in Tables 1$-$8 are implemented in the R programming language based on 500 replications.
\subsection{Evaluation of Estimation Efficiency}
\subsubsection{Scenario 1: Fixed \texorpdfstring{$N_B$}{} and Varying Batch Size \texorpdfstring{$n_b$}{}}

We first consider a scenario with fixed $N_B$ and varying batch size $n_b$. Specifically, we generate $B$ data blocks consisting of $N_B = 10^5$ independent observations, where each data batch has $n_b$ observations. In Tables 1-4, we  present the results for $\bbeta_1$ (intercept) and $\bbeta_2$ ($\bbeta_i$'s are similar to $\bbeta_2$ and omitted, $i=3,4,5$). From Tables 1-4 we can see that the proposed renewable estimator performs comparably with the traditional LPRE estimator based on the entire data.  Moreover,  the bias of the CUEE is much smaller than that of the CEE estimator, while its SSE is much larger than that of the CEE as $n_b$ decreases to 50. The coverage probabilities of the CUEE estimator are mostly dropped below 90\% as $n_b$ decreases to 50. This confirms that if the condition $B=O(n_b^j)$, $j<1/3$ is not satisfied, the CEE and CUEE methods do not have valid asymptotic distributions for inference. In contrast, the proposed method achieves valid and efficient statistical efficiency, and its coverage probability is around 95\% under the chosen settings.

%In contrast, the CEE method is unstable and its coverage probabilities are the smallest as the batch size $n_b$ decreases to 200.

\subsubsection{Scenario 2: Fixed Batch Size \texorpdfstring{$n_b$}{} and Varying \texorpdfstring{$B$}{}}
Now we consider another scenario with fixed batch size $n_b$ and varying $B$. For convenience, we fix batch size $n_b = 100$, and $N_B$ is varying from $10^3$ to $10^5$. In Tables 5-8, we  present the estimation results for $\bbeta_1$ (intercept) and $\bbeta_2$, respectively ($\bbeta_i$'s are similar to $\bbeta_2$ and omitted, $i=3,4,5$). It is clear to see that the SSE and ESE of all estimators decrease as $B$ increases, which verifies the consistency of the four estimators. Moreover, it is shown that the four estimators are unbiased, the SSE is close to the ESE, and the coverage probabilities are satisfactory when $B=10$ and $100$. However, as $B$ increases to $10^3$, the SSE and ESE of the CUEE are significantly larger than that of other competitors. Although the ESE of the CEE is smaller than the proposed method, its coverage probability is much smaller compared with the full data LPRE when $\log(\epsilon)$ follows $N(0, 1)$.  On the contrary, the proposed renewable method always exhibits the similar performance to the traditional LPRE estimator as $B$ increases from 10 to $10^3$, which confirms the stability and effectiveness of the proposed renewable estimator.

%However, as $B$ increases to $10^2$, the CEE's coverage probabilities are the smallest compared with other three competitors.

\subsection{ Evaluation of Computation Efficiency}
To assess the computation efficiency, we report the CPU time (in seconds) for the LPRE, CEE, CUEE and the proposed method. We consider the following two scenarios: (a) varying $B$ (fixed $N_B=10^7$); (b) varying $N_b$ (fixed  $B=10^3$). All computations are carried out on a laptop running R programming language with 16GB random-access memory (RAM). The results are the mean CPU time of ten replications. Tables 9 and 10 report C.Time and R.Time for two scenarios mentioned above with Case 1 and $\log(\epsilon)\sim N(0, 1)$. From Table 9, we can see that the proposed method is always much faster than the other three competitors. Moreover, the C.Time of the proposed method is less than 12 seconds, while the traditional LPRE requires more than 500 seconds when the number of batches $B$ increases to $10^4$. This fast computation of the proposed method does not lose statistical efficiency. In addition, the CUEE takes more computing time compared with the CEE. The main reason is that the CEE estimator does not require an additional step to calculate the intermediary estimator $\cbeta_b$ and the aggregated matrix $\sum_{k=1}^{b}\mathbf{S}_k(\mathcal{D}_k,\cbeta_k)$. As shown in Table 10, compared with the other three competitors, our proposed renewable learning method has significant computational advantages both in R.Time and C.Time as $N_B$ increases to $10^8$.

\section{Real Data Analysis}

\subsection{The Bike Sharing Data}
In this section, we apply our proposed method to the bike sharing dataset, which is publicly available at {\it http://archive.ics.uci.edu/ml/datasets/Bike+Sharing+Dataset}. The streaming datasets arrive monthly during the 24-month period
from January 2011 to December 2012, where $B= 24$ and $N_b=17379$. We consider four covariates : a binary variable  ``workingday" ($X_1$) to indicate whether a certain day is a working day or not(1 = working day; 0 = non-working day), three continuous variables:  temperature ($X_2$), humidity ($X_3$) and windspeed ($X_4$). The square of the number of bikes rented hourly is used as the response. Similar to the simulation studies, we also compare our proposed method with the CEE, CUEE and traditional LPRE method. We report the estimated parameters, standard errors and $p$-values in Table 11. As we can see, all $p$-values are sufficiently small ($\ll0.05$), which indicates that all covariates are significant towards the response. It is seen from Table 11 that the number of rented bikes in non-working days is more than that of working days. The temperature and windspeed have positive influences on the number of rented bikes, and the humidity has a negative effect. Additionally, the CUEE has slight larger standard errors than those of the CEE and our proposed renewable method, which is in line with the simulation results.
\subsection{The Electric Power Consumption Data}
We apply our proposed method to an electric power consumption dataset, which contains 2,049,280 completed measurements for a house located at Sceaux between December 2006 and November 2010. We consider the scenario where the data arrive monthly during the 48-month period with $B=48$ data batches. The data is publicly available at {\it http://archive.ics.uci.edu/ml/datasets/Individual+household+electric+power+consumption}. For the analysis, the minute-averaged current intensity (in ampere) is used as the response. We consider three covariates: active electrical energy  in the kitchen ($X_1$, in watt-hour), active electrical energy in the  laundry room ($X_2$, in watt-hour), and active electrical energy for an electric water-heater and an air-conditioner ($X_3$, in watt-hour). All covariates are centered and scaled with mean 0 and variance 1. Similar to Section 6.1, we present the estimated coefficients, standard errors and $p$-values in Table 12. As shown in Table 12, three online updating estimates are unbiased and all standard errors are significantly small. In addition, all $p$-values are small enough to indicate that each covariate is significant towards the response. It is not surprising that there is more electric current through the water-heater and air-conditioner than through the laundry room and kitchen. This is because the power of the water-heater and air-conditioner is relatively large.
\section{Concluding Remarks}
In this paper, we proposed a renewable estimation method for the multiplicative regression model in the streaming data environment. The consistency and asymptotic normality of the renewable estimator were established. The simulation studies showed that the proposed renewable estimator was desirable compared with the CEE and CUEE estimator. In addition, the proposed estimator was asymptotically equivalent with the traditional LPRE estimator based on the entire data available. Two real data examples illustrated the effectiveness of our proposed renewable method.

There are several important topics to investigate further in the future. First, the paper mainly considered the problem of parameter estimation and statistical inference for multiplicative regression with streaming data. However, the online variable selection was not considered in the online updating context. This problem is interesting especially when accessing historical data is limited. Second, it is worth to extend our proposed renewable method to other survival models, such as the Cox model of \cite{cox1972regression} and the additive hazards model in \cite{lin1994semiparametric}.

\section*{Acknowledgements}
Sun's research is partially supported by the National Natural Science Foundation of China (12171463).

\section*{Appendix}

\renewcommand{\theequation}{A.\arabic{equation}} % (S. equation)
\setcounter{equation}{0}

\begin{lemma}\label{le1}
If the conditions (C.1) and (C.2) hold, we have
\begin{align} \label{A1}
\mathbf{S}_b(\mathcal{D}_b,\bbeta_t)= O_p(\sqrt{n_b}).
\end{align}
\end{lemma}
{\bf Proof}.
Direct calculation yields that
\begin{align} \label{A2}
&\mathbb{E}\{\mathbf{S}_b(\mathcal{D}_b,\bbeta_t)\}=0,
\end{align}
and
\begin{align} \label{A3}
&Var\{\mathbf{S}_b(\mathcal{D}_b,\bbeta_t)\}=\mathbb{E}\left[\sum_{i\in\mathcal{D}_b}
\left\{Y_{i}^{-1}\exp(\bbeta_t^T\mathbf{X}_{i})-Y_{i} \exp(-\bbeta_t^T\mathbf{X}_{i})\right\}^2\mathbf{X}_{i}\mathbf{X}_{i}\right]\nonumber\\
&~~~~~~~~~~~~~~~~~~~~~~\leq\mathbb{E}\left[\sum_{i\in\mathcal{D}_b}
\left\{Y_{i}^{-1}\exp(\bbeta_t^T\mathbf{X}_{i})+Y_{i} \exp(-\bbeta_t^T\mathbf{X}_{i})\right\}^2\mathbf{X}_{i}\mathbf{X}_{i}\right]\nonumber\\
&~~~~~~~~~~~~~~~~~~~~~~=O_P(n_b),
\end{align}
where the last equality is due to the condition (C.3).
Combining (\ref{A2}), (\ref{A3}) and Chebyshev's inequality, (\ref{A1}) follows. This ends the proof.
%%%%%%%%%%%

\vspace{1cm}
\noindent
{\bf Proof of Theorem \ref{th1}.} We define a function
\begin{align*}
f_b(\bbeta)=-\frac{1}{N_b}\sum_{k=1}^{b-1}\mathbf{Q}_k(\mathcal{D}_k,\tbeta_k)(\bbeta-\tbeta_{b-1})+\frac{1}{N_b}\mathbf{S}_b
(\mathcal{D}_b,\bbeta).
\end{align*}
According to (\ref{Eq3.2}), the renewable estimator $\tbeta_2$ satisfies
\begin{align}\label{A4}
f_2(\tbeta_2)=0.
\end{align}
%Due to the fact that $\tbeta_{1}-\bbeta_t = o_p(1)$. By the conditions (C.3) and (C.4), we can derive that
%\begin{align} \label{A5}
%&\mathbf{Q}_1(\mathcal{D}_1,\tbeta_1)\leq\mathbf{Q}_1(\mathcal{D}_1,\bbeta_t)+ M(\mathcal{D}_1)\|\tbeta_1-\bbeta_t\|\nonumber\\
%&~~~~~~~~~~~~~~=O_p(n_1)+O_p(n_1) o_p(1)\nonumber\\
%&~~~~~~~~~~~~~~=O_p(n_1).
%\end{align}
Based on (\ref{A1}) and the fact that $\tbeta_{1}-\bbeta_t = o_p(1)$,  it can be verified that
\begin{align}\label{A6}
f_2(\bbeta_t)=\frac{1}{N_2}\mathbf{Q}_1(\mathcal{D}_1,\tbeta_1)(\tbeta_1-\bbeta_t)+\frac{1}{N_2}\mathbf{S}_2
(\mathcal{D}_2,\bbeta_t)=o_p(1),
\end{align}
as $N_2\rightarrow\infty$. By taking a difference between equations (\ref{A6}) and (\ref{A4}), we can get
\begin{align}\label{A7}
&f_2(\bbeta_t)-f_2(\tbeta_2)=\frac{1}{N_2}\mathbf{Q}_1(\mathcal{D}_1,\tbeta_1)(\tbeta_2-\bbeta_t)-\frac{1}{N_2}
\mathbf{S}_2(\mathcal{D}_2,\tbeta_2)+\frac{1}{N_2}\mathbf{S}_2(\mathcal{D}_2,\bbeta_t)\nonumber\\
&~~~~~~~~~~~~~~~~~~~~=o_p(1).
\end{align}
In addition, by taking the first-order Taylor expansion of $\mathbf{S}_2(\mathcal{D}_2,\tbeta_2)$ around $\bbeta_t$, we obtain
\begin{align}\label{A8}
\mathbf{S}_2(\mathcal{D}_2,\tbeta_2)=\mathbf{S}_2(\mathcal{D}_2,\bbeta_t)-\mathbf{Q}_2(\mathcal{D}_2,\bbeta_t)(\tbeta_2-\bbeta_t)
+O_p\left(\|\tbeta_2-\bbeta_t\|\right).
\end{align}
In view of (\ref{A7}) and (\ref{A8}), we can derive
\begin{align*}
&f_2(\bbeta_t)-f_2(\tbeta_2)=\frac{1}{N_2}\left\{\mathbf{Q}_1(\mathcal{D}_1,\tbeta_1)+
\mathbf{Q}_2(\mathcal{D}_2,\bbeta_t)\right\}(\tbeta_2-\bbeta_t)+O_p\left(\frac{1}{N_2}\|\tbeta_2-\bbeta_t\|\right)\nonumber\\
&~~~~~~~~~~~~~~~~~~~~=o_p(1).
\end{align*}
Under the condition (C.2), we know that $\frac{1}{N_2}\left\{\mathbf{Q}_1(\mathcal{D}_1,\tbeta_1)+
\mathbf{Q}_2(\mathcal{D}_2,\bbeta_t)\right\}$ is positive-definite. Therefore, we have $\tbeta_2-\bbeta_t\stackrel{P}{\longrightarrow}0$ as $N_2\rightarrow\infty$.

After some similar derivations, it can be easily shown that
\begin{align}\label{A9}
\tbeta_k-\bbeta_t=o_p(1),
\end{align}
for $k=1,\ldots,b-1$. According to (\ref{Eq3.4}), the renewable estimator $\tbeta_b$ satisfies
\begin{align}\label{A10}
f_b(\tbeta_b)=0.
\end{align}
%Under the conditions (C.3) and (C.4), it can be shown that
%\begin{align} \label{A10}
%&\mathbf{Q}_k(\mathcal{D}_k,\tbeta_k)\leq\mathbf{Q}_k(\mathcal{D}_k,\bbeta_t)+ M(\mathcal{D}_k)\|\tbeta_k-\bbeta_t\|\nonumber\\
%&~~~~~~~~~~~~~~~=O_p(n_k)+O_p(n_k) o_p(1)\nonumber\\
%&~~~~~~~~~~~~~~~=O_p(n_k),
%\end{align}
%for $k=1,\ldots,b-1$.
From (\ref{A1}) and (\ref{A9}), we obtain
\begin{align}\label{A11}
f_b(\bbeta_t)=\frac{1}{N_b}\sum_{k=1}^{b-1}\mathbf{Q}_k(\mathcal{D}_k,\tbeta_k)(\tbeta_{b-1}-\bbeta_t)+\frac{1}{N_b}\mathbf{S}_b
(\mathcal{D}_b,\bbeta_t)=o_p(1),
\end{align}
as $N_b\rightarrow\infty$. By taking a difference between (\ref{A11}) and (\ref{A10}), we get
\begin{align}\label{A12}
&f_b(\bbeta_t)-f_b(\tbeta_b)=\frac{1}{N_b}\sum_{k=1}^{b-1}\mathbf{Q}_k(\mathcal{D}_k,\tbeta_k)(\tbeta_b-\bbeta_t)-\frac{1}{N_b}
\mathbf{S}_b(\mathcal{D}_b,\tbeta_b)+\frac{1}{N_b}\mathbf{S}_b(\mathcal{D}_b,\bbeta_t)\nonumber\\
&~~~~~~~~~~~~~~~~~~~=o_p(1).
\end{align}
Similar to (\ref{A8}), taking the first-order Taylor expansion of $\mathbf{S}_b(\mathcal{D}_b,\tbeta_b)$ around $\bbeta_t$, we have
\begin{align}\label{A13}
\mathbf{S}_b(\mathcal{D}_b,\tbeta_b)=\mathbf{S}_b(\mathcal{D}_b,\bbeta_t)-\mathbf{Q}_b(\mathcal{D}_b,\bbeta_t)(\tbeta_b-\bbeta_t)
+O_p\left(\|\tbeta_b-\bbeta_t\|\right).
\end{align}
It follows from (\ref{A12}) and (\ref{A13}) that
\begin{align}\label{A14}
&f_b(\bbeta_t)-f_b(\tbeta_b)=\frac{1}{N_b}\left\{\sum_{k=1}^{b-1}\mathbf{Q}_k(\mathcal{D}_k,\tbeta_k)+
\mathbf{Q}_b(\mathcal{D}_b,\bbeta_t)\right\}(\tbeta_b-\bbeta_t)+O_p\left(\frac{1}{N_b}\|\tbeta_b-\bbeta_t\|\right)\nonumber\\
&~~~~~~~~~~~~~~~~~~~=o_p(1).
\end{align}
Since $\frac{1}{N_b}\left\{\sum_{k=1}^{b-1}\mathbf{Q}_k(\mathcal{D}_k,\tbeta_k)+
\mathbf{Q}_b(\mathcal{D}_b,\bbeta_t)\right\}$ is positive-definite, we have $\tbeta_b-\bbeta_t\stackrel{P}{\longrightarrow}0$ as $N_b\rightarrow\infty$. This completes the proof.
%%%%%%%%%%%%%%%%%%%%%%%

\vspace{1cm}
\noindent
{\bf Proof of Theorem \ref{th2}.}
By \cite{chen2016least}, we know that the traditional LPRE $\tbeta_1$ satisfies $\frac{1}{N_1}\mathbf{S}_1(\mathcal{D}_1,\tbeta_1)=0$ and
\begin{align*}
\sqrt{N_1}(\tbeta_1-\bbeta_t)\stackrel{{d}}{\longrightarrow} N(\mathbf{0},\mathbb{G}^{-1}(\bbeta_t)),
\end{align*}
as $N_1\rightarrow\infty$. Besides, using the Taylor expansion method, the score function has the following expression:
\begin{align*}
\frac{1}{N_1}\mathbf{S}_1(\mathcal{D}_1,\bbeta_t)=\frac{1}{N_1}\mathbf{S}_1(\mathcal{D}_1,\tbeta_1)+\frac{1}{N_1}\mathbf{Q}_1(\mathcal{D}_1,\tbeta_1)(\tbeta_1-\bbeta_t)
+O_p\left(\frac{1}{N_1}\|\bbeta_t-\tbeta_1\|^2\right).
\end{align*}
Notice that $\frac{1}{N_1}\mathbf{S}_1(\mathcal{D}_1,\tbeta_1)=0$, we can get
\begin{align}\label{A15}
\frac{1}{N_1}\mathbf{S}_1(\mathcal{D}_1,\bbeta_t)=\frac{1}{N_1}\mathbf{Q}_1(\mathcal{D}_1,\tbeta_1)(\tbeta_1-\bbeta_t)
+o_p(1).
\end{align}
From (\ref{A4}), (\ref{A8}), (\ref{A15}) and  Theorem \ref{th1}, we know that
\begin{align*}
\frac{1}{N_2}\left\{\mathbf{S}_1(\mathcal{D}_1,\bbeta_t)+\mathbf{S}_2(\mathcal{D}_2,\bbeta_t)\right\}=
\frac{1}{N_2}\left\{\mathbf{Q}_1(\mathcal{D}_1,\tbeta_1)+\mathbf{Q}_2(\mathcal{D}_2,\tbeta_2)\right\}(\tbeta_{2}-\bbeta_t)+o_p(1).
\end{align*}
Similarly, at the $(b-1)$th data batch, we have
\begin{align}\label{A16}
\frac{1}{N_{b-1}}\sum_{k=1}^{b-1}\mathbf{S}_k(\mathcal{D}_k,\bbeta_t)=\frac{1}{N_{b-1}}\sum_{k=1}^{b-1}
\mathbf{Q}_k(\mathcal{D}_k,\tbeta_k)(\tbeta_{b-1}-\bbeta_t)+o_p(1).
\end{align}
Based on (\ref{A10}), (\ref{A11}) and (\ref{A14}), it can be  verified that
\begin{align}\label{A17}
&-\frac{1}{N_b}\left\{\sum_{k=1}^{b-1}\mathbf{Q}_k(\mathcal{D}_k,\tbeta_k)+\mathbf{Q}_b(\mathcal{D}_b,\bbeta_t)\right\}
(\tbeta_b-\bbeta_t)+\frac{1}{N_b}\sum_{k=1}^{b-1}\mathbf{Q}_k(\mathcal{D}_k,\tbeta_k)(\tbeta_{b-1}-\bbeta_t)\nonumber\\
&+\frac{1}{N_b}\mathbf{S}_b(\mathcal{D}_b,\bbeta_t)+o_p(1)=0.
\end{align}
Combining (\ref{A16}) and (\ref{A17}), it follows that
\begin{align*}
\frac{1}{N_b}\sum_{k=1}^{b}\mathbf{S}_k(\mathcal{D}_k,\bbeta_t)-\frac{1}{N_b}\left\{\sum_{k=1}^{b-1}
\mathbf{Q}_k(\mathcal{D}_k,\tbeta_k)+\mathbf{Q}_b(\mathcal{D}_b,\bbeta_t)\right\}(\tbeta_b-\bbeta_t)+o_p(1)=0.
\end{align*}
According to Theorem \ref{th1}, all $\tbeta_k$'s are consistent, $k=1,\ldots,b$. By the Continuous Mapping Theorem (Theorem 5.1 in \citealt{billingsley1968convergence}), we can deduce that
\begin{align*}
&\frac{1}{N_{b}}\sum_{k=1}^{b}\mathbf{S}_k(\mathcal{D}_k,\bbeta_t)-\frac{1}{N_b}\sum_{k=1}^{b}
\mathbf{Q}_k(\mathcal{D}_k,\bbeta_t)(\tbeta_b-\bbeta_t)+o_p(1)=0.
\end{align*}
By the condition (C.2), we have
\begin{align*}
\sqrt{N_b}(\tbeta_b-\bbeta_t)=\left\{\frac{1}{N_b}\sum_{k=1}^{b}\mathbf{Q}_k(\mathcal{D}_k,\bbeta_t)\right\}^{-1}\left\{\frac{1}{\sqrt{N_b}}
\sum_{k=1}^{b}\mathbf{S}_k(\mathcal{D}_k,\bbeta_t)\right\}+o_p(1).
\end{align*}
Note that $\mathbf{Q}_k(\mathcal{D}_k,\tbeta_k)$ and $\mathbf{C}_k(\mathcal{D}_k,\tbeta_k)$ are consistent to $\mathbf{Q}_k(\mathcal{D}_k,\bbeta_t)$ and $\mathbf{C}_k(\mathcal{D}_k,\bbeta_t)$, respectively. Then we have
\begin{align*}
\frac{1}{N_b}\left\{\tilde{\mathbf{Q}}_b^T \tilde{\mathbf{C}}_b^{-1}\tilde{\mathbf{Q}}_b\right\}\stackrel{P}{\longrightarrow}\mathbb{G}(\bbeta_t).
\end{align*}
From the central limit theorem and Slutsky's theorem, we get
\begin{align*}
\sqrt{N_b}(\tbeta_b-\bbeta_t)\stackrel{{d}}{\longrightarrow} N(\mathbf{0},\mathbb{G}^{-1}(\bbeta_t)),
\end{align*}
where $\mathbb{G}(\bbeta_t)=\mathbb{Q}^T(\bbeta_t)\mathbb{C}^{-1}(\bbeta_t)\mathbb{Q}(\bbeta_t)$. This ends the proof.

\bibliographystyle{natbib}
\bibliography{reference}

\begin{thebibliography}{}

\bibitem[Billingsley(1968)]{billingsley1968convergence}
Billingsley, P. (1968).
\newblock \emph{Convergence of Probability Measures}.
\newblock Wiley, New York.

\bibitem[Chen \emph{et~al.}(2010)Chen, Guo, Lin, and Ying]{chen2010least}
Chen, K., Guo, S., Lin, Y., and Ying, Z. (2010).
\newblock Least absolute relative error estimation.
\newblock \emph{Journal of the American Statistical Association} \textbf{105},
  1104--1112.

\bibitem[Chen \emph{et~al.}(2016)Chen, Lin, Wang, and Ying]{chen2016least}
Chen, K., Lin, Y., Wang, Z., and Ying, Z. (2016).
\newblock Least product relative error estimation.
\newblock \emph{Journal of Multivariate Analysis} \textbf{144}, 91--98.

\bibitem[Cox(1972)]{cox1972regression}
Cox, D.~R. (1972).
\newblock Regression models and life-tables.
\newblock \emph{Journal of the Royal Statistical Society: Series B
  (Methodological)} \textbf{34}, 187--202.

\bibitem[Hector \emph{et~al.}(2021)Hector, Luo, and Song]{hector2021parallel}
Hector, E.~C., Luo, L., and Song, P. X.-K. (2021).
\newblock Parallel-and-stream accelerator for computationally fast supervised
  learning.
\newblock \emph{arXiv:2111.00032} .

\bibitem[Jin \emph{et~al.}(2003)Jin, Lin, Wei, and Ying]{jin2003rank}
Jin, Z., Lin, D., Wei, L., and Ying, Z. (2003).
\newblock Rank-based inference for the accelerated failure time model.
\newblock \emph{Biometrika} \textbf{90}, 341--353.

\bibitem[Lee \emph{et~al.}(2020)Lee, Wang, and Schifano]{lee2020online}
Lee, J., Wang, H., and Schifano, E.~D. (2020).
\newblock Online updating method to correct for measurement error in big data
  streams.
\newblock \emph{Computational Statistics and Data Analysis} \textbf{149},
  106976.

\bibitem[Li \emph{et~al.}(2014)Li, Lin, Zhou, and Zhou]{li2014empirical}
Li, Z., Lin, Y., Zhou, G., and Zhou, W. (2014).
\newblock Empirical likelihood for least absolute relative error regression.
\newblock \emph{Test} \textbf{23}, 86--99.

\bibitem[Lin and Ying(1994)]{lin1994semiparametric}
Lin, D.~Y. and Ying, Z. (1994).
\newblock Semiparametric analysis of the additive risk model.
\newblock \emph{Biometrika} \textbf{81}, 61--71.

\bibitem[Lin \emph{et~al.}(2020)Lin, Li, and Lu]{lin2020unified}
Lin, L., Li, W., and Lu, J. (2020).
\newblock Unified rules of renewable weighted sums for various online updating
  estimations.
\newblock \emph{arXiv:2008.08824} .

\bibitem[Lin \emph{et~al.}(2021)Lin, Lu, and Li]{lu2021online}
Lin, L., Lu, J., and Li, W. (2021).
\newblock Online updating statistics for heterogenous updating regressions via
  homogenization techniques.
\newblock \emph{arXiv:2106.12370} .

\bibitem[Luo \emph{et~al.}(2021)Luo, Han, Lin, and Huang]{luo2021statistical}
Luo, L., Han, R., Lin, Y., and Huang, J. (2021).
\newblock Statistical inference in high-dimensional generalized linear models
  with streaming data.
\newblock \emph{arXiv:2108.04437} .

\bibitem[Luo and Song(2020)]{luo2020renewable}
Luo, L. and Song, P. X.-K. (2020).
\newblock Renewable estimation and incremental inference in generalized linear
  models with streaming data sets.
\newblock \emph{Journal of the Royal Statistical Society: Series B (Statistical
  Methodology)} \textbf{82}, 69--97.

\bibitem[Luo and Song(2021)]{luo2021multivariate}
Luo, L. and Song, P. X.-K. (2021).
\newblock Multivariate online regression analysis with heterogeneous streaming
  data.
\newblock \emph{Canadian Journal of Statistics} DOI:{
  \href{https://doi.org/10.1002/cjs.11667}{10.1002/cjs.11667}}.

\bibitem[Luo \emph{et~al.}(2022)Luo, Zhou, and Song]{luo2022real}
Luo, L., Zhou, L., and Song, P. X.-K. (2022).
\newblock Real-time regression analysis of streaming clustered data with
  possible abnormal data batches.
\newblock \emph{Journal of the American Statistical Association} DOI:{
  \href{https://doi.org/10.1080/01621459.2022.2026778}{10.1080/01621459.2022.2026778}}.

\bibitem[Schifano \emph{et~al.}(2016)Schifano, Wu, Wang, Yan, and
  Chen]{schifano2016online}
Schifano, E.~D., Wu, J., Wang, C., Yan, J., and Chen, M.-H. (2016).
\newblock Online updating of statistical inference in the big data setting.
\newblock \emph{Technometrics} \textbf{58}, 393--403.

\bibitem[Shi and Luo(2021)]{shi2021online}
Shi, X. and Luo, L. (2021).
\newblock Online causal inference with application to near real-time
  post-market vaccine safety surveillance.
\newblock \emph{arXiv:2111.13775} .

\bibitem[Teekens and Koerts(1972)]{teekens1972some}
Teekens, R. and Koerts, J. (1972).
\newblock Some statistical implications of the log transformation of
  multiplicative models.
\newblock \emph{Econometrica} \textbf{40}, 793--819.

\bibitem[Wang \emph{et~al.}(2022)Wang, Wang, and Li]{wang2022renewable}
Wang, K., Wang, H., and Li, S. (2022).
\newblock Renewable quantile regression for streaming datasets.
\newblock \emph{Knowledge-Based Systems} \textbf{235}, 107675.

\bibitem[Wei(1992)]{wei1992accelerated}
Wei, L.-J. (1992).
\newblock The accelerated failure time model: a useful alternative to the cox
  regression model in survival analysis.
\newblock \emph{Statistics in Medicine} \textbf{11}, 1871--1879.

\bibitem[Wu \emph{et~al.}(2021)Wu, Chen, Schifano, and Yan]{wu2021online}
Wu, J., Chen, M.-H., Schifano, E.~D., and Yan, J. (2021).
\newblock Online updating of survival analysis.
\newblock \emph{Journal of Computational and Graphical Statistics} \textbf{30},
  1209--1223.

\bibitem[Xia \emph{et~al.}(2016)Xia, Liu, and Yang]{xia2016regularized}
Xia, X., Liu, Z., and Yang, H. (2016).
\newblock Regularized estimation for the least absolute relative error models
  with a diverging number of covariates.
\newblock \emph{Computational Statistics and Data Analysis} \textbf{96},
  104--119.

\bibitem[Xue \emph{et~al.}(2020)Xue, Wang, Yan, and Schifano]{xue2020online}
Xue, Y., Wang, H., Yan, J., and Schifano, E.~D. (2020).
\newblock An online updating approach for testing the proportional hazards
  assumption with streams of survival data.
\newblock \emph{Biometrics} \textbf{76}, 171--182.

\end{thebibliography}

\clearpage
\begin{sidewaystable}[ht]
\begin{center}
 {{\bf Table 1}. Simulation results for the estimator ${\tbeta_1}$ with varying batch size $n_b$ and $log(\epsilon )\sim N(0,1)$.}\\
\vspace{0.1in}
\small
\begin{tabular}{ccccccccccccccccccc}
\hline
&\multirow{2}{*}{case 1}  &LPRE   &  &  \multicolumn{3}{c}{CEE}       &  &    \multicolumn{3}{c}{CUEE}    &  & \multicolumn{3}{c}{Renew}\\
\cline{5-7}\cline{9-11}\cline{13-15}
&                     &$n_b=10^5$ &  &$n_b=1000$ &$n_b=200$ &$n_b=50$ &  &$n_b=1000$ &$n_b=200$ &$n_b=50$ &  &$n_b=1000$ &$n_b=200$ &$n_b=50$\\
&$BIAS\times 10^{-4}$ &1.60       &  &1.59       &1.62      &1.67     &  &1.58       &1.50      &0.96     &  &1.60       &1.60      &1.60\\
&$SSE\times 10^{-3}$  &3.49       &  &3.49       &3.48      &3.46     &  &3.49       &3.47      &4.18     &  &3.49       &3.49      &3.49\\
&$ESE\times 10^{-3}$  &3.43       &  &3.39       &3.29      &3.05     &  &3.43       &3.44      &3.49     &  &3.42       &3.42      &3.42\\
&CP                   &0.954      &  &0.952      &0.948     &0.920    &  &0.958      &0.958     &0.928    &  &0.954      &0.954     &0.954\\
\hline
&\multirow{2}{*}{case 2}  &LPRE   &  &  \multicolumn{3}{c}{CEE}       &  &    \multicolumn{3}{c}{CUEE}    &  & \multicolumn{3}{c}{Renew}\\
\cline{5-7}\cline{9-11}\cline{13-15}
&                     &$n_b=10^5$ &  &$n_b=1000$ &$n_b=200$ &$n_b=50$ &  &$n_b=1000$ &$n_b=200$ &$n_b=50$ &  &$n_b=1000$ &$n_b=200$ &$n_b=50$\\
&$BIAS\times 10^{-4}$ &$-$2.34    &  &$-$2.34    &$-$2.34   &$-$2.32  &  &$-$2.34    &$-$1.35   &$-$0.29  &  &$-$2.34    &$-$2.34   &$-$2.34\\
&$SSE\times 10^{-3}$  &3.50       &  &3.50       &3.50      &3.47     &  &3.50       &3.84      &9.52     &  &3.50       &3.50      &3.50\\
&$ESE\times 10^{-3}$  &3.43       &  &3.39       &3.29      &3.05     &  &3.43       &3.46      &3.53     &  &3.43       &3.43      &3.43\\
&CP                   &0.944      &  &0.942      &0.928     &0.908    &  &0.944      &0.928     &0.728    &  &0.944      &0.944     &0.944\\
\hline
&\multirow{2}{*}{case 3}  &LPRE   &  &  \multicolumn{3}{c}{CEE}       &  &    \multicolumn{3}{c}{CUEE}    &  & \multicolumn{3}{c}{Renew}\\
\cline{5-7}\cline{9-11}\cline{13-15}
&                     &$n_b=10^5$ &  &$n_b=1000$ &$n_b=200$ &$n_b=50$ &  &$n_b=1000$ &$n_b=200$ &$n_b=50$ &  &$n_b=1000$ &$n_b=200$ &$n_b=50$\\
&$BIAS\times 10^{-4}$ &2.30       &  &2.31       &2.33      &2.42     &  &2.29       &2.16      &1.02     &  &2.28       &2.28      &2.31\\
&$SSE\times 10^{-3}$  &3.30       &  &3.30       &3.29      &3.29     &  &3.30       &3.33      &3.86     &  &3.30       &3.30      &3.30\\
&$ESE\times 10^{-3}$  &3.43       &  &3.40       &3.30      &3.07     &  &3.43       &3.44      &3.49     &  &3.43       &3.43      &3.43\\
&CP                   &0.940      &  &0.940      &0.936     &0.914    &  &0.940      &0.948     &0.924    &  &0.940      &0.940     &0.940\\
\hline
&\multirow{2}{*}{case 4}  &LPRE   &  &  \multicolumn{3}{c}{CEE}       &  &    \multicolumn{3}{c}{CUEE}    &  & \multicolumn{3}{c}{Renew}\\
\cline{5-7}\cline{9-11}\cline{13-15}
&                     &$n_b=10^5$ &  &$n_b=1000$ &$n_b=200$ &$n_b=50$ &  &$n_b=1000$ &$n_b=200$ &$n_b=50$ &  &$n_b=1000$ &$n_b=200$ &$n_b=50$\\
&$BIAS\times 10^{-4}$ &9.25       &  &9.28       &9.44      &9.60     &  &9.23       &9.21      &8.27     &  &9.26       &9.27      &9.28\\
&$SSE\times 10^{-3}$  &7.72       &  &7.72       &7.69      &7.59     &  &7.71       &7.72      &10.92    &  &7.72       &7.72      &7.72\\
&$ESE\times 10^{-3}$  &7.67       &  &7.56       &7.32      &6.79     &  &7.68       &7.72      &7.88     &  &7.66       &7.66      &7.66\\
&CP                   &0.948      &  &0.946      &0.938     &0.914    &  &0.950      &0.950     &0.866    &  &0.948      &0.948     &0.948\\
\hline
\end{tabular}
\end{center}
\end{sidewaystable}

\clearpage
\begin{sidewaystable}[ht]
\begin{center}
 {{\bf Table 2}. Simulation results for the estimator ${\tbeta_2}$ with varying batch size $n_b$ and $log(\epsilon )\sim N(0,1)$.}\\
\vspace{0.1in}
\small
\begin{tabular}{ccccccccccccccccccc}
\hline
&\multirow{2}{*}{case 1}  &LPRE   &  &  \multicolumn{3}{c}{CEE}       &  &    \multicolumn{3}{c}{CUEE}    &  & \multicolumn{3}{c}{Renew}\\
\cline{5-7}\cline{9-11}\cline{13-15}
&                     &$n_b=10^5$ &  &$n_b=1000$ &$n_b=200$ &$n_b=50$ &  &$n_b=1000$ &$n_b=200$ &$n_b=50$ &  &$n_b=1000$ &$n_b=200$ &$n_b=50$\\
&$BIAS\times 10^{-4}$ &$-$1.15    &  &$-$1.13    &$-$1.04   &$-$0.74  &  &$-$1.07    &$-$0.78   &$-$0.08  &  &$-$1.15    &$-$1.15   &$-$1.15\\
&$SSE\times 10^{-3}$  &4.10       &  &4.10       &4.08      &4.04     &  &4.10       &4.13      &5.51     &  &4.10       &4.10      &4.10\\
&$ESE\times 10^{-3}$  &3.96       &  &3.90       &3.75      &3.43     &  &3.96       &3.98      &4.05     &  &3.96       &3.95      &3.95\\
&CP                   &0.930      &  &0.930      &0.922     &0.896    &  &0.930      &0.926     &0.872    &  &0.930      &0.930     &0.930\\
\hline
&\multirow{2}{*}{case 2}  &LPRE   &  &  \multicolumn{3}{c}{CEE}       &  &    \multicolumn{3}{c}{CUEE}    &  & \multicolumn{3}{c}{Renew}\\
\cline{5-7}\cline{9-11}\cline{13-15}
&                     &$n_b=10^5$ &  &$n_b=1000$ &$n_b=200$ &$n_b=50$ &  &$n_b=1000$ &$n_b=200$ &$n_b=50$ &  &$n_b=1000$ &$n_b=200$ &$n_b=50$\\
&$BIAS\times 10^{-4}$ &1.18       &  &1.20       &1.24      &1.30     &  &1.12       &0.56      &$-$0.99  &  &1.19       &1.19      &1.20\\
&$SSE\times 10^{-3}$  &3.65       &  &3.66       &3.65      &3.62     &  &3.66       &3.83      &7.12     &  &3.65       &3.65      &3.65\\
&$ESE\times 10^{-3}$  &3.73       &  &3.68       &3.54      &3.24     &  &3.74       &3.77      &3.87     &  &3.73       &3.73      &3.73\\
&CP                   &0.952      &  &0.950      &0.942     &0.916    &  &0.952      &0.948     &0.822    &  &0.952      &0.952     &0.952\\
\hline
&\multirow{2}{*}{case 3}  &LPRE   &  &  \multicolumn{3}{c}{CEE}       &  &    \multicolumn{3}{c}{CUEE}    &  & \multicolumn{3}{c}{Renew}\\
\cline{5-7}\cline{9-11}\cline{13-15}
&                     &$n_b=10^5$ &  &$n_b=1000$ &$n_b=200$ &$n_b=50$ &  &$n_b=1000$ &$n_b=200$ &$n_b=50$ &  &$n_b=1000$ &$n_b=200$ &$n_b=50$\\
&$BIAS\times 10^{-4}$ &0.76       &  &0.79       &0.82      &1.41     &  &0.59       &$-$0.38   &$-$1.28  &  &1.07       &1.09      &0.77\\
&$SSE\times 10^{-3}$  &3.00       &  &2.99       &2.98      &2.93     &  &3.00       &3.22      &9.46     &  &3.08       &3.08      &2.99\\
&$ESE\times 10^{-3}$  &3.07       &  &3.00       &2.85      &2.57     &  &3.07       &3.12      &3.39     &  &3.06       &3.06      &3.06\\
&CP                   &0.964      &  &0.960      &0.946     &0.910    &  &0.964      &0.958     &0.736    &  &0.962      &0.962     &0.964\\
\hline
&\multirow{2}{*}{case 4}  &LPRE   &  &  \multicolumn{3}{c}{CEE}       &  &    \multicolumn{3}{c}{CUEE}    &  & \multicolumn{3}{c}{Renew}\\
\cline{5-7}\cline{9-11}\cline{13-15}
&                     &$n_b=10^5$ &  &$n_b=1000$ &$n_b=200$ &$n_b=50$ &  &$n_b=1000$ &$n_b=200$ &$n_b=50$ &  &$n_b=1000$ &$n_b=200$ &$n_b=50$\\
&$BIAS\times 10^{-4}$ &$-$3.77    &  &$-$3.76    &$-$3.74   &$-$4.21  &  &$-$3.74    &$-$3.45   &1.97     &  &$-$3.77    &$-$3.78   &$-$3.79\\
&$SSE\times 10^{-3}$  &6.96       &  &6.96       &6.93      &6.85     &  &6.97       &7.14      &12.82    &  &6.96       &6.96      &6.95\\
&$ESE\times 10^{-3}$  &6.86       &  &6.71       &6.39      &5.77     &  &6.87       &6.93      &7.14     &  &6.85       &6.85      &6.85\\
&CP                   &0.954      &  &0.948      &0.940     &0.908    &  &0.954      &0.948     &0.852    &  &0.954      &0.954     &0.956\\
\hline
\end{tabular}
\end{center}
\end{sidewaystable}

\clearpage
\begin{sidewaystable}[ht]
\begin{center}
 {{\bf Table 3}. Simulation results for the estimator ${\tbeta_1}$ with varying batch size $n_b$ and $log(\epsilon)\sim Uniform(-2,2)$.}\\
\vspace{0.1in}
\small
\begin{tabular}{ccccccccccccccccccc}
\hline
&\multirow{2}{*}{case 1}  &LPRE   &  &  \multicolumn{3}{c}{CEE}       &  &    \multicolumn{3}{c}{CUEE}    &  & \multicolumn{3}{c}{Renew}\\
\cline{5-7}\cline{9-11}\cline{13-15}
&                     &$n_b=10^5$ &  &$n_b=1000$ &$n_b=200$ &$n_b=50$ &  &$n_b=1000$ &$n_b=200$ &$n_b=50$ &  &$n_b=1000$ &$n_b=200$ &$n_b=50$\\
&$BIAS\times 10^{-4}$ &1.58       &  &1.57       &1.57      &1.52     &  &1.56       &1.47      &0.35     &  &1.58       &1.58      &1.58\\
&$SSE\times 10^{-3}$  &3.06       &  &3.06       &3.06      &3.07     &  &3.06       &3.05      &3.63     &  &3.06       &3.06      &3.06\\
&$ESE\times 10^{-3}$  &2.98       &  &2.97       &2.97      &2.94     &  &2.98       &2.99      &3.02     &  &2.98       &2.98      &2.98\\
&CP                   &0.948      &  &0.948      &0.946     &0.942    &  &0.948      &0.954     &0.890    &  &0.948      &0.948     &0.948\\
\hline
&\multirow{2}{*}{case 2}  &LPRE   &  &  \multicolumn{3}{c}{CEE}       &  &    \multicolumn{3}{c}{CUEE}    &  & \multicolumn{3}{c}{Renew}\\
\cline{5-7}\cline{9-11}\cline{13-15}
&                     &$n_b=10^5$ &  &$n_b=1000$ &$n_b=200$ &$n_b=50$ &  &$n_b=1000$ &$n_b=200$ &$n_b=50$ &  &$n_b=1000$ &$n_b=200$ &$n_b=50$\\
&$BIAS\times 10^{-4}$ &$-$3.59    &  &$-$3.59    &$-$3.61   &$-$3.67  &  &$-$3.56    &$-$3.57   &0.59     &  &$-$3.59    &$-$3.59   &$-$3.59\\
&$SSE\times 10^{-3}$  &2.95       &  &2.95       &2.95      &2.96     &  &2.95       &3.16      &8.49     &  &2.95       &2.95      &2.95\\
&$ESE\times 10^{-3}$  &2.98       &  &2.97       &2.97      &2.94     &  &2.98       &3.00      &3.06     &  &2.98       &2.98      &2.98\\
&CP                   &0.948      &  &0.948      &0.946     &0.940    &  &0.948      &0.928     &0.716    &  &0.948      &0.948     &0.948\\
\hline
&\multirow{2}{*}{case 3}  &LPRE   &  &  \multicolumn{3}{c}{CEE}       &  &    \multicolumn{3}{c}{CUEE}    &  & \multicolumn{3}{c}{Renew}\\
\cline{5-7}\cline{9-11}\cline{13-15}
&                     &$n_b=10^5$ &  &$n_b=1000$ &$n_b=200$ &$n_b=50$ &  &$n_b=1000$ &$n_b=200$ &$n_b=50$ &  &$n_b=1000$ &$n_b=200$ &$n_b=50$\\
&$BIAS\times 10^{-4}$ &$-$2.62    &  &$-$2.63    &$-$2.66   &$-$2.78  &  &$-$2.62    &$-$2.53   &$-$2.49  &  &$-$2.68    &$-$2.63   &$-$2.63\\
&$SSE\times 10^{-3}$  &2.84       &  &2.84       &2.84      &2.85     &  &2.84       &2.85      &3.22     &  &2.84       &2.84      &2.84\\
&$ESE\times 10^{-3}$  &2.98       &  &2.97       &2.97      &2.94     &  &2.98       &2.99      &3.03     &  &2.98       &2.98      &2.98\\
&CP                   &0.956      &  &0.956      &0.954     &0.956    &  &0.956      &0.956     &0.922    &  &0.956      &0.956     &0.956\\
\hline
&\multirow{2}{*}{case 4}  &LPRE   &  &  \multicolumn{3}{c}{CEE}       &  &    \multicolumn{3}{c}{CUEE}    &  & \multicolumn{3}{c}{Renew}\\
\cline{5-7}\cline{9-11}\cline{13-15}
&                     &$n_b=10^5$ &  &$n_b=1000$ &$n_b=200$ &$n_b=50$ &  &$n_b=1000$ &$n_b=200$ &$n_b=50$ &  &$n_b=1000$ &$n_b=200$ &$n_b=50$\\
&$BIAS\times 10^{-4}$ &$-$5.19    &  &$-$5.21    &$-$5.28   &$-$5.12  &  &$-$5.11    &$-$5.07   &$-$3.83  &  &$-$5.19    &$-$5.20   &$-$5.20\\
&$SSE\times 10^{-3}$  &6.69       &  &6.69       &6.67      &6.73     &  &6.69       &6.80      &10.42    &  &6.69       &6.69      &6.69\\
&$ESE\times 10^{-3}$  &6.65       &  &6.65       &6.63      &6.52     &  &6.66       &6.70      &6.83     &  &6.65       &6.65      &6.65\\
&CP                   &0.940      &  &0.940      &0.940     &0.934    &  &0.940      &0.940     &0.830    &  &0.940      &0.940     &0.940\\
\hline
\end{tabular}
\end{center}
\end{sidewaystable}

\clearpage
\begin{sidewaystable}[ht]
\begin{center}
 {{\bf Table 4}. Simulation results for the estimator ${\tbeta_2}$ with varying batch size $n_b$ and $log(\epsilon)\sim Uniform(-2,2)$.}\\
\vspace{0.1in}
\small
\begin{tabular}{ccccccccccccccccccc}
\hline
&\multirow{2}{*}{case 1}  &LPRE   &  &  \multicolumn{3}{c}{CEE}       &  &    \multicolumn{3}{c}{CUEE}    &  & \multicolumn{3}{c}{Renew}\\
\cline{5-7}\cline{9-11}\cline{13-15}
&                     &$n_b=10^5$ &  &$n_b=1000$ &$n_b=200$ &$n_b=50$ &  &$n_b=1000$ &$n_b=200$ &$n_b=50$ &  &$n_b=1000$ &$n_b=200$ &$n_b=50$\\
&$BIAS\times 10^{-4}$ &1.02       &  &1.02       &0.95      &0.90     &  &1.00       &0.94      &$-$0.06  &  &1.02       &1.02      &1.02\\
&$SSE\times 10^{-3}$  &3.55       &  &3.55       &3.55      &3.58     &  &3.55       &3.56      &4.55     &  &3.55       &3.55      &3.55\\
&$ESE\times 10^{-3}$  &3.44       &  &3.43       &3.43      &3.37     &  &3.44       &3.46      &3.51     &  &3.44       &3.44      &3.44\\
&CP                   &0.948      &  &0.948      &0.948     &0.944    &  &0.948      &0.946     &0.902    &  &0.948      &0.948     &0.948\\
\hline
&\multirow{2}{*}{case 2}  &LPRE   &  &  \multicolumn{3}{c}{CEE}       &  &    \multicolumn{3}{c}{CUEE}    &  & \multicolumn{3}{c}{Renew}\\
\cline{5-7}\cline{9-11}\cline{13-15}
&                     &$n_b=10^5$ &  &$n_b=1000$ &$n_b=200$ &$n_b=50$ &  &$n_b=1000$ &$n_b=200$ &$n_b=50$ &  &$n_b=1000$ &$n_b=200$ &$n_b=50$\\
&$BIAS\times 10^{-4}$ &1.03       &  &1.03       &0.99      &1.13     &  &1.02       &0.97      &$-$0.87  &  &1.03       &1.03      &1.03\\
&$SSE\times 10^{-3}$  &3.29       &  &3.29       &3.29      &3.32     &  &3.29       &3.37      &5.43     &  &3.29       &3.29      &3.29\\
&$ESE\times 10^{-3}$  &3.24       &  &3.24       &3.23      &3.18     &  &3.24       &3.27      &3.35     &  &3.24       &3.24      &3.24\\
&CP                   &0.948      &  &0.948      &0.948     &0.938    &  &0.950      &0.948     &0.820    &  &0.948      &0.948     &0.948\\
\hline
&\multirow{2}{*}{case 3}  &LPRE   &  &  \multicolumn{3}{c}{CEE}       &  &    \multicolumn{3}{c}{CUEE}    &  & \multicolumn{3}{c}{Renew}\\
\cline{5-7}\cline{9-11}\cline{13-15}
&                     &$n_b=10^5$ &  &$n_b=1000$ &$n_b=200$ &$n_b=50$ &  &$n_b=1000$ &$n_b=200$ &$n_b=50$ &  &$n_b=1000$ &$n_b=200$ &$n_b=50$\\
&$BIAS\times 10^{-4}$ &0.57       &  &0.57       &0.50      &0.57     &  &0.53       &0.47      &0.42     &  &0.57       &0.60      &0.57\\
&$SSE\times 10^{-3}$  &2.65       &  &2.65       &2.66      &2.69     &  &2.65       &2.85      &7.10     &  &2.65       &2.65      &2.65\\
&$ESE\times 10^{-3}$  &2.66       &  &2.66       &2.64      &2.56     &  &2.67       &2.71      &2.93     &  &2.66       &2.66      &2.66\\
&CP                   &0.942      &  &0.940      &0.940     &0.926    &  &0.942      &0.936     &0.758    &  &0.942      &0.942     &0.942\\
\hline
&\multirow{2}{*}{case 4}  &LPRE   &  &  \multicolumn{3}{c}{CEE}       &  &    \multicolumn{3}{c}{CUEE}    &  & \multicolumn{3}{c}{Renew}\\
\cline{5-7}\cline{9-11}\cline{13-15}
&                     &$n_b=10^5$ &  &$n_b=1000$ &$n_b=200$ &$n_b=50$ &  &$n_b=1000$ &$n_b=200$ &$n_b=50$ &  &$n_b=1000$ &$n_b=200$ &$n_b=50$\\
&$BIAS\times 10^{-4}$ &5.64       &  &5.71       &5.87      &5.79     &  &5.62       &5.21      &3.41     &  &5.64       &5.64      &5.64 \\
&$SSE\times 10^{-3}$  &5.84       &  &5.84       &5.83      &5.91     &  &5.84       &5.98      &10.55    &  &5.84       &5.84      &5.84\\
&$ESE\times 10^{-3}$  &5.95       &  &5.95       &5.91      &5.73     &  &5.96       &6.01      &6.20     &  &5.95       &5.95      &5.95\\
&CP                   &0.948      &  &0.948      &0.948     &0.934    &  &0.948      &0.944     &0.812    &  &0.948      &0.948     &0.948\\
\hline
\end{tabular}
\end{center}
\end{sidewaystable}

\clearpage
\begin{sidewaystable}[ht]
\begin{center}
 {{\bf Table 5}. Simulation results for the estimator ${\tbeta_1}$ with varying $B$ and $log(\epsilon )\sim N(0,1)$.}\\
\vspace{0.1in}
\small
\resizebox{\textwidth}{80mm}{
\begin{tabular}{ccccccccccccccccccc}
\hline
&\multirow{2}{*}{case 1}  &\multicolumn{3}{c}{LPRE} & & \multicolumn{3}{c}{CEE} &  & \multicolumn{3}{c}{CUEE}  &  & \multicolumn{3}{c}{Renew}\\
\cline{3-5}\cline{7-9}\cline{11-13}\cline{15-17}
&                      &$B=10$ &$B=100$ &$B=10^3$ &  &$B=10$ &$B=100$ &$B=10^3$ &  &$B=10$ &$B=100$ &$B=10^3$&  &$B=10$ &$B=100$ &$B=10^3$\\
&$BIAS\times10^{-4}$  &$-$4.03&$-$0.09&1.31   &       &$-$4.67&$-$0.12&1.31   &      &$-$4.01&0.01   &1.28   &      &$-$4.11&$-$0.09&1.31 \\
&$SSE\times10^{-3}$   &34.33  &10.60  &3.39   &       &34.34  &10.56  &3.39   &      &34.32  &10.61  &3.46   &      &34.34  &10.59  &3.39 \\
&$ESE\times10^{-3}$   &33.50  &10.80  &3.43   &       &31.67  &10.09  &3.20   &      &33.20  &10.85  &3.46   &      &33.07  &10.76  &3.42 \\
&CP                   &0.948  &0.960  &0.942  &       &0.932  &0.938  &0.926  &      &0.948  &0.958  &0.942  &      &0.948  &0.958  &0.942\\
\hline
&\multirow{2}{*}{case 2}  &\multicolumn{3}{c}{LPRE} & & \multicolumn{3}{c}{CEE} &  & \multicolumn{3}{c}{CUEE}  &  & \multicolumn{3}{c}{Renew}\\
\cline{3-5}\cline{7-9}\cline{11-13}\cline{15-17}
&                      &$B=10$ &$B=100$ &$B=10^3$ &  &$B=10$ &$B=100$ &$B=10^3$ &  &$B=10$ &$B=100$ &$B=10^3$&  &$B=10$ &$B=100$ &$B=10^3$\\
&$BIAS\times10^{-4}$  &$-$4.94&$-$0.02&1.29   &       &$-$4.90&0.12   &1.29   &      &$-$4.70&$-$0.01&1.29   &     &$-$5.20&$-$0.04&1.29 \\
&$SSE\times10^{-3}$   &33.91  &10.58  &3.39   &       &33.80  &10.55  &3.38   &      &33.88  &10.60  &3.48   &     &33.90  &10.58  &3.39 \\
&$ESE\times10^{-3}$   &33.52  &10.80  &3.43   &       &31.69  &10.09  &3.19   &      &33.21  &10.85  &3.46   &     &33.06  &10.76  &3.42\\
&CP                   &0.948  &0.956  &0.942  &       &0.936  &0.938  &0.928  &      &0.948  &0.960  &0.936  &     &0.948  &0.958  &0.942\\
\hline
&\multirow{2}{*}{case 3}  &\multicolumn{3}{c}{LPRE} & & \multicolumn{3}{c}{CEE} &  & \multicolumn{3}{c}{CUEE}  &  & \multicolumn{3}{c}{Renew}\\
\cline{3-5}\cline{7-9}\cline{11-13}\cline{15-17}
&                      &$B=10$ &$B=100$ &$B=10^3$ &  &$B=10$ &$B=100$ &$B=10^3$ &  &$B=10$ &$B=100$ &$B=10^3$&  &$B=10$ &$B=100$ &$B=10^3$\\
&$BIAS\times10^{-4}$  &$-$0.40&0.02   &2.05   &       &$-$1.23&$-$0.16&2.17   &      &$-$0.59&0.14   &2.08   &      &$-$1.00&0.01   &2.04\\
&$SSE\times10^{-3}$   &35.31  &10.59  &3.53   &       &35.33  &10.58  &3.52   &      &35.33  &10.64  &3.68   &      &35.31  &10.59  & 3.53 \\
&$ESE\times10^{-3}$   &33.81  &10.80  &3.43   &       &32.03  &10.13  &3.21   &      &33.53  &10.85  &3.46   &      &33.39  &10.76  & 3.42\\
&CP                   &0.926  &0.960  &0.944  &       &0.912  &0.938  &0.928  &      &0.920  &0.958  &0.932  &      &0.920  &0.958  & 0.944\\
\hline
&\multirow{2}{*}{case 4}  &\multicolumn{3}{c}{LPRE} & & \multicolumn{3}{c}{CEE} &  & \multicolumn{3}{c}{CUEE}  &  & \multicolumn{3}{c}{Renew}\\
\cline{3-5}\cline{7-9}\cline{11-13}\cline{15-17}
&                      &$B=10$ &$B=100$ &$B=10^3$ &  &$B=10$ &$B=100$ &$B=10^3$ &  &$B=10$ &$B=100$ &$B=10^3$&  &$B=10$ &$B=100$ &$B=10^3$\\
&$BIAS\times10^{-4}$  &17.73   &5.61  &4.45   &       &19.53  &6.11   &4.88   &      &18.12  &5.27   &1.93   &      &18.24  &5.91   & 4.44\\
&$SSE\times10^{-3}$   &77.34   &23.79  &7.33   &      &77.16  &23.71  &7.29   &      &77.47  &24.39  &8.86   &      &77.31  &23.81  & 7.33 \\
&$ESE\times10^{-3}$   &74.97   &24.14  &7.66   &      &70.86  &22.44  &7.10   &      &74.45  &24.41  &7.78   &      &73.97  &24.04  & 7.66\\
&CP                   &0.950   &0.950  &0.964  &      &0.938  &0.930  &0.944  &      &0.948  &0.946  &0.936  &      &0.946  &0.948  & 0.964\\
\hline
\end{tabular}}
\end{center}
\end{sidewaystable}

\clearpage
\begin{sidewaystable}[ht]
\begin{center}
 {{\bf Table 6}. Simulation results for the estimator ${\tbeta_2}$ with varying $B$ and $log(\epsilon )\sim N(0,1)$.}\\
\vspace{0.1in}
\small
\resizebox{\textwidth}{80mm}{
\begin{tabular}{ccccccccccccccccccc}
\hline
&\multirow{2}{*}{case 1}  &\multicolumn{3}{c}{LPRE} & & \multicolumn{3}{c}{CEE} &  & \multicolumn{3}{c}{CUEE}  &  & \multicolumn{3}{c}{Renew}\\
\cline{3-5}\cline{7-9}\cline{11-13}\cline{15-17}
&                      &$B=10$ &$B=100$ &$B=10^3$ &  &$B=10$ &$B=100$ &$B=10^3$ &  &$B=10$ &$B=100$ &$B=10^3$&  &$B=10$ &$B=100$ &$B=10^3$\\
&$BIAS\times10^{-4}$  &10.49  &0.92   &1.82   &       &11.14  &1.13   &1.93   &      &10.06  &0.96   &1.61   &    &10.52  &0.85   &1.82 \\
&$SSE\times10^{-3}$   &38.29  &12.24  &3.88   &       &38.01  &12.18  &3.86   &      &38.23  &12.32  &4.27   &    &38.21  &12.23  &3.88\\
&$ESE\times10^{-3}$   &38.58  &12.46  &3.96   &       &35.98  &11.44  &3.62   &      &38.15  &12.53  &4.00   &    &37.92  &12.39  &3.95\\
&CP                   &0.942  &0.964  &0.964  &       &0.922  &0.946  &0.946  &      &0.938  &0.958  &0.944  &    &0.936  &0.962  &0.964\\
\hline
&\multirow{2}{*}{case 2}  &\multicolumn{3}{c}{LPRE} & & \multicolumn{3}{c}{CEE} &  & \multicolumn{3}{c}{CUEE}  &  & \multicolumn{3}{c}{Renew}\\
\cline{3-5}\cline{7-9}\cline{11-13}\cline{15-17}
&                      &$B=10$ &$B=100$ &$B=10^3$ &  &$B=10$ &$B=100$ &$B=10^3$ &  &$B=10$ &$B=100$ &$B=10^3$&  &$B=10$ &$B=100$ &$B=10^3$\\
&$BIAS\times10^{-4}$  &$-$7.40&$-$7.56&$-$3.17&       &$-$6.40&$-$7.35&$-$3.15&      &$-$6.21&$-$6.94&$-$2.55&      &$-$7.08&$-$7.54&$-$3.16\\
&$SSE\times10^{-3}$   &34.19  &11.74  &3.81   &       &34.05  &11.68  &3.81   &      &34.20  &11.89  &4.11   &      &34.17  &11.73  &3.81 \\
&$ESE\times10^{-3}$   &36.11  &11.73  &3.73   &       &33.76  &10.79  &3.42   &      &35.71  &11.82  &3.78   &      &35.50  &11.67  &3.73\\
&CP                   &0.966  &0.942  &0.948  &       &0.954  &0.924  &0.920  &      &0.960  &0.946  &0.936  &      &0.960  &0.942  &0.946\\
\hline
&\multirow{2}{*}{case 3}  &\multicolumn{3}{c}{LPRE} & & \multicolumn{3}{c}{CEE} &  & \multicolumn{3}{c}{CUEE}  &  & \multicolumn{3}{c}{Renew}\\
\cline{3-5}\cline{7-9}\cline{11-13}\cline{15-17}
&                      &$B=10$ &$B=100$ &$B=10^3$ &  &$B=10$ &$B=100$ &$B=10^3$ &  &$B=10$ &$B=100$ &$B=10^3$&  &$B=10$ &$B=100$ &$B=10^3$\\
&$BIAS\times10^{-4}$  &$-$0.84&$-$1.72&$-$3.20&       &$-$0.74&$-$2.55&$-$3.46&      &$-$0.17&$-$1.95&$-$2.02&      &$-$0.16&$-$1.68&$-$2.93\\
&$SSE\times10^{-3}$   &31.15  &9.02   &3.01   &       &31.13  &8.94   &3.00   &      &31.20  &9.46   &5.63   &      &31.21  &9.01   & 3.00 \\
&$ESE\times10^{-3}$   &29.48  &9.63   &3.06   &       &27.28  &8.66   &2.74   &      &29.25  &9.80   &3.23   &      &28.95  &9.56   & 3.06\\
&CP                   &0.942  &0.950  &0.950  &       &0.916  &0.930  &0.926  &      &0.940  &0.948  &0.862  &      &0.940  &0.950  & 0.950\\
\hline
&\multirow{2}{*}{case 4}  &\multicolumn{3}{c}{LPRE} & & \multicolumn{3}{c}{CEE} &  & \multicolumn{3}{c}{CUEE}  &  & \multicolumn{3}{c}{Renew}\\
\cline{3-5}\cline{7-9}\cline{11-13}\cline{15-17}
&                      &$B=10$ &$B=100$ &$B=10^3$ &  &$B=10$ &$B=100$ &$B=10^3$ &  &$B=10$ &$B=100$ &$B=10^3$&  &$B=10$ &$B=100$ &$B=10^3$\\
&$BIAS\times10^{-4}$  &$-$21.85&$-$2.93&$-$1.31&     &$-$22.59&$-$2.58&$-$1.84&      &$-$22.18&$-$0.64&1.52   &      &$-$22.76&$-$3.03&$-$1.32\\
&$SSE\times10^{-3}$   &65.35   &20.43  &6.60   &     &65.02   &20.40  &6.52   &      &65.60   &20.89  &8.85   &      &65.30   &20.45  & 6.61 \\
&$ESE\times10^{-3}$   &65.50   &21.56  &6.84   &     &60.65   &19.40  &6.13   &      &64.78   &21.91  &6.99   &      &64.24   &21.42  & 6.83\\
&CP                   &0.950   &0.964  &0.966  &     &0.924   &0.944  &0.944  &      &0.948   &0.970  &0.922  &      &0.942   &0.964  & 0.966\\
\hline
\end{tabular}}
\end{center}
\end{sidewaystable}

\clearpage
\begin{sidewaystable}[ht]
\begin{center}
 {{\bf Table 7}. Simulation results for the estimator ${\tbeta_1}$ with varying $B$ and $log(\epsilon)\sim Uniform(-2,2)$.}\\
\vspace{0.1in}
\small
\resizebox{\textwidth}{80mm}{
\begin{tabular}{ccccccccccccccccccc}
\hline
&\multirow{2}{*}{case 1}  &\multicolumn{3}{c}{LPRE} & & \multicolumn{3}{c}{CEE} &  & \multicolumn{3}{c}{CUEE}  &  & \multicolumn{3}{c}{Renew}\\
\cline{3-5}\cline{7-9}\cline{11-13}\cline{15-17}
&                      &$B=10$ &$B=100$ &$B=10^3$ &  &$B=10$ &$B=100$ &$B=10^3$ &  &$B=10$ &$B=100$ &$B=10^3$&  &$B=10$ &$B=100$ &$B=10^3$\\
&$BIAS\times10^{-4}$  &1.52   &1.96   &$-$4.02&       &2.47   &1.98   &$-$4.06&      &1.55   &1.81   &$-$3.81&      &1.64   &1.96   &$-$4.02\\
&$SSE\times10^{-3}$   &30.12  &8.81   &2.74   &       &30.14  &8.81   &2.74   &      &30.13  &8.86   &2.83   &      &30.13  &8.81   &2.74 \\
&$ESE\times10^{-3}$   &29.77  &9.41   &2.98   &       &29.66  &9.38   &2.96   &      &29.84  &9.47   &3.00   &      &29.75  &9.41   &2.98 \\
&CP                   &0.958  &0.964  &0.964  &       &0.956  &0.960  &0.964  &      &0.958  &0.962  &0.954  &      &0.956  &0.962  &0.964\\
\hline
&\multirow{2}{*}{case 2}  &\multicolumn{3}{c}{LPRE} & & \multicolumn{3}{c}{CEE} &  & \multicolumn{3}{c}{CUEE}  &  & \multicolumn{3}{c}{Renew}\\
\cline{3-5}\cline{7-9}\cline{11-13}\cline{15-17}
&                      &$B=10$ &$B=100$ &$B=10^3$ &  &$B=10$ &$B=100$ &$B=10^3$ &  &$B=10$ &$B=100$ &$B=10^3$&  &$B=10$ &$B=100$ &$B=10^3$\\
&$BIAS\times10^{-4}$  &1.87   &1.97   &$-$0.75&       &2.31   &2.14   &$-$0.76&      &1.73   &1.85   &$-$0.51&     &1.83   &1.97   &$-$0.75\\
&$SSE\times10^{-3}$   &30.18  &8.82   &3.03   &       &30.18  &8.82   &3.04   &      &30.16  &8.88   &3.18   &     &30.16  &8.82   &3.03 \\
&$ESE\times10^{-3}$   &29.76  &9.41   &2.98   &       &29.65  &9.38   &2.96   &      &29.84  &9.48   &3.00   &     &29.74  &9.41   &2.98\\
&CP                   &0.956  &0.962  &0.952  &       &0.954  &0.958  &0.948  &      &0.958  &0.962  &0.936  &     &0.956  &0.962  &0.952\\
\hline
&\multirow{2}{*}{case 3}  &\multicolumn{3}{c}{LPRE} & & \multicolumn{3}{c}{CEE} &  & \multicolumn{3}{c}{CUEE}  &  & \multicolumn{3}{c}{Renew}\\
\cline{3-5}\cline{7-9}\cline{11-13}\cline{15-17}
&                      &$B=10$ &$B=100$ &$B=10^3$ &  &$B=10$ &$B=100$ &$B=10^3$ &  &$B=10$ &$B=100$ &$B=10^3$&  &$B=10$ &$B=100$ &$B=10^3$\\
&$BIAS\times10^{-4}$  &$-$12.06&3.47   &1.89   &     &$-$12.24&3.41   &1.95   &      &$-$11.89&2.76   &1.52   &      &$-$12.26&3.45   &1.89\\
&$SSE\times10^{-3}$   &31.33   &9.17   &2.96   &     &31.35   &9.18   &2.97   &      &31.33   &9.18   &3.01   &      &31.32   &9.17   & 2.96 \\
&$ESE\times10^{-3}$   &29.78   &9.41   &2.98   &     &29.66   &9.37   &2.96   &      &29.86   &9.48   &3.00   &      &29.76   &9.41   & 2.98\\
&CP                   &0.948   &0.956  &0.942  &     &0.944   &0.954  &0.946  &      &0.948   &0.960  &0.940  &      &0.948   &0.956  & 0.942\\
\hline
&\multirow{2}{*}{case 4}  &\multicolumn{3}{c}{LPRE} & & \multicolumn{3}{c}{CEE} &  & \multicolumn{3}{c}{CUEE}  &  & \multicolumn{3}{c}{Renew}\\
\cline{3-5}\cline{7-9}\cline{11-13}\cline{15-17}
&                      &$B=10$ &$B=100$ &$B=10^3$ &  &$B=10$ &$B=100$ &$B=10^3$ &  &$B=10$ &$B=100$ &$B=10^3$&  &$B=10$ &$B=100$ &$B=10^3$\\
&$BIAS\times10^{-4}$  &55.07   &$-$8.32&$-$2.22&      &56.47  &$-$7.80&$-$2.23&      &54.70  &$-$7.43&$-$1.47&      &55.13  &$-$8.36&$-$2.22\\
&$SSE\times10^{-3}$   &65.06   &20.50  &6.68   &      &65.30  &20.56  &6.72   &      &65.17  &20.57  &7.44   &      &65.16  &20.50  & 6.68 \\
&$ESE\times10^{-3}$   &66.50   &21.04  &6.65   &      &65.96  &20.86  &6.60   &      &66.72  &21.27  &6.74   &      &66.40  &21.03  & 6.65\\
&CP                   &0.938   &0.958  &0.944  &      &0.944  &0.960  &0.946  &      &0.938  &0.960  &0.912  &      &0.938  &0.958  & 0.944\\
\hline
\end{tabular}}
\end{center}
\end{sidewaystable}

\clearpage
\begin{sidewaystable}[ht]
\begin{center}
 {{\bf Table 8}. Simulation results for the estimator ${\tbeta_2}$ with varying $B$ and $log(\epsilon)\sim Uniform(-2,2)$.}\\
\vspace{0.1in}
\small
\resizebox{\textwidth}{80mm}{
\begin{tabular}{ccccccccccccccccccc}
\hline
&\multirow{2}{*}{case 1}  &\multicolumn{3}{c}{LPRE} & & \multicolumn{3}{c}{CEE} &  & \multicolumn{3}{c}{CUEE}  &  & \multicolumn{3}{c}{Renew}\\
\cline{3-5}\cline{7-9}\cline{11-13}\cline{15-17}
&                      &$B=10$ &$B=100$ &$B=10^3$ &  &$B=10$ &$B=100$ &$B=10^3$ &  &$B=10$ &$B=100$ &$B=10^3$&  &$B=10$ &$B=100$ &$B=10^3$\\
&$BIAS\times10^{-4}$  &$-$21.12&4.00   &$-$1.10&     &$-$22.51&3.90   &$-$1.06&      &$-$21.15&3.44   &$-$0.48&    &$-$21.15&4.02   &$-$1.10\\
&$SSE\times10^{-3}$   &33.64   &10.52  &3.32   &     &33.72   &10.54  &3.31   &      &33.67   &10.54  &3.63   &    &33.68   &10.52  &3.32\\
&$ESE\times10^{-3}$   &34.39   &10.87  &3.44   &     &34.18   &10.80  &3.42   &      &34.52   &10.97  &3.48   &    &34.37   &10.87  &3.44\\
&CP                   &0.962   &0.966  &0.958  &     &0.954   &0.964  &0.954  &      &0.962   &0.962  &0.936  &    &0.962   &0.966  &0.958\\
\hline
&\multirow{2}{*}{case 2}  &\multicolumn{3}{c}{LPRE} & & \multicolumn{3}{c}{CEE} &  & \multicolumn{3}{c}{CUEE}  &  & \multicolumn{3}{c}{Renew}\\
\cline{3-5}\cline{7-9}\cline{11-13}\cline{15-17}
&                      &$B=10$ &$B=100$ &$B=10^3$ &  &$B=10$ &$B=100$ &$B=10^3$ &  &$B=10$ &$B=100$ &$B=10^3$&  &$B=10$ &$B=100$ &$B=10^3$\\
&$BIAS\times10^{-4}$  &19.84  &$-$2.10&$-$0.76&       &20.55  &$-$2.13&$-$0.82&      &19.85  &$-$1.32&$-$0.66&      &19.71  &$-$2.13&$-$0.76\\
&$SSE\times10^{-3}$   &32.21  &10.26  &3.26   &       &32.23  &10.27  &3.25   &      &32.24  &10.30  &3.39   &      &32.21  &10.26  &3.26 \\
&$ESE\times10^{-3}$   &32.47  &10.24  &3.24   &       &32.28  &10.18  &3.22   &      &32.59  &10.35  &3.27   &      &32.44  &10.24  &3.24\\
&CP                   &0.956  &0.948  &0.942  &       &0.952  &0.944  &0.944  &      &0.952  &0.960  &0.938  &      &0.954  &0.946  &0.942\\
\hline
&\multirow{2}{*}{case 3}  &\multicolumn{3}{c}{LPRE} & & \multicolumn{3}{c}{CEE} &  & \multicolumn{3}{c}{CUEE}  &  & \multicolumn{3}{c}{Renew}\\
\cline{3-5}\cline{7-9}\cline{11-13}\cline{15-17}
&                      &$B=10$ &$B=100$ &$B=10^3$ &  &$B=10$ &$B=100$ &$B=10^3$ &  &$B=10$ &$B=100$ &$B=10^3$&  &$B=10$ &$B=100$ &$B=10^3$\\
&$BIAS\times10^{-4}$  &0.11   &$-$0.74&$-$1.29&       &2.04   &$-$0.63&$-$1.46&      &0.09   &$-$0.15&$-$0.61&      &0.47   &$-$0.80&$-$1.29\\
&$SSE\times10^{-3}$   &26.50  &8.29   &2.76   &       &26.52  &8.37   &2.80   &      &26.42  &8.74   &3.77   &      &26.45  &8.29   & 2.76 \\
&$ESE\times10^{-3}$   &26.69  &8.42   &2.66   &       &26.30  &8.28   &2.62   &      &26.85  &8.65   &2.77   &      &26.63  &8.42   & 2.66\\
&CP                   &0.940  &0.960  &0.944  &       &0.936  &0.958  &0.944  &      &0.942  &0.950  &0.912  &      &0.936  &0.960  & 0.946\\
\hline
&\multirow{2}{*}{case 4}  &\multicolumn{3}{c}{LPRE} & & \multicolumn{3}{c}{CEE} &  & \multicolumn{3}{c}{CUEE}  &  & \multicolumn{3}{c}{Renew}\\
\cline{3-5}\cline{7-9}\cline{11-13}\cline{15-17}
&                      &$B=10$ &$B=100$ &$B=10^3$ &  &$B=10$ &$B=100$ &$B=10^3$ &  &$B=10$ &$B=100$ &$B=10^3$&  &$B=10$ &$B=100$ &$B=10^3$\\
&$BIAS\times10^{-4}$  &$-$41.34&6.31   &2.04   &     &$-$42.09&6.16   &1.95   &      &$-$41.60&4.98   &0.68   &      &$-$40.81&6.34   &2.04\\
&$SSE\times10^{-3}$   &59.29   &19.41  &6.10   &     &59.47   &19.46  &6.16   &      &59.42   &19.67  &7.20   &      &59.23   &19.40  & 6.10 \\
&$ESE\times10^{-3}$   &59.60   &18.81  &5.95   &     &58.69   &18.53  &5.86   &      &60.00   &19.16  &6.08   &      &59.46   &18.81  & 5.95\\
&CP                   &0.946   &0.944  &0.942  &     &0.938   &0.938  &0.940  &      &0.948   &0.938  &0.904  &      &0.944   &0.944  & 0.942\\
\hline
\end{tabular}}
\end{center}
\end{sidewaystable}

\clearpage
\begin{table}[ht]
\begin{center}
 {{\bf Table 9}. The computing time in seconds for Case 1 with varying $B$ and fixed the cumulative sample size $N_b=10^7$ .}\\
\vspace{0.1in}
\small
\begin{tabular}{cccccccccccccccc}
\hline
&  & \multicolumn{2}{c}{  LPRE} &  & \multicolumn{2}{c}{ CEE}& &\multicolumn{2}{c}{ CUEE} &  & \multicolumn{2}{c}{ Renew}\\
\cline{3-4}\cline{6-7}\cline{9-10}\cline{12-13}
&   B         &  C.Time    &  R.Time     &    &  C.Time    &  R.Time     &  &  C.Time    &  R.Time     &  &  C.Time    &  R.Time \\
\hline
&  10         &  12.689    &  10.090     &    &  11.841    &  9.984      &  &  14.743    &  12.846     &  &  9.751     &  7.914  \\
&  $10^2$     &  17.922    &  10.195     &    &  11.446    &  9.544      &  &  14.425    &  12.490     &  &  9.075     &  7.155  \\
&  $10^3$     &  48.489    &  9.461      &    &  10.642    &  8.270      &  &  13.101    &  11.075     &  &  8.526     &  6.436  \\
&  $10^4$     &  511.851   &  9.489      &    &  14.463    &  10.326     &  &  17.330    &  13.172     &  &  11.439    &  7.473  \\
\hline
\end{tabular}
\end{center}
\end{table}

\begin{table}[ht]
\begin{center}
 {{\bf Table 10}. The computing time in seconds for Case 1 with varying the cumulative sample size $N_b$ and fixed  $B=10^3$ .}\\
\vspace{0.1in}
\small
\begin{tabular}{cccccccccccccccc}
\hline
&  & \multicolumn{2}{c}{  LPRE} &  & \multicolumn{2}{c}{ CEE}& &\multicolumn{2}{c}{ CUEE} &  & \multicolumn{2}{c}{ Renew}\\
\cline{3-4}\cline{6-7}\cline{9-10}\cline{12-13}
&   N       &  C.Time   &  R.Time     &    &  C.Time    &  R.Time     &  &  C.Time    &  R.Time     &  &  C.Time    &  R.Time \\
\hline
&  $10^6$          &  4.360    &  0.957      &    &  1.471     &  1.077      &  &  1.767     &  1.344      &  &  1.172     &  0.758  \\
&  $5\times10^6$   &  35.748   &  5.100      &    &  5.501     &  4.344      &  &  6.819     &  5.679      &  &  4.362     &  3.209  \\
&  $10^7$          &  58.028   &  9.859      &    &  10.889    &  8.731      &  &  13.514    &  11.418     &  &  8.707     &  6.615  \\
&  $5\times10^7$   &  385.513  &  52.000     &    &  56.306    &  46.454     &  &  69.927    &  60.128     &  &  44.129    &  34.441  \\
&  $10^8$          &  978.540  &  164.629    &    &  121.692   &  101.209    &  &  148.298   &  127.412    &  &  93.962    &  73.506  \\
\hline
\end{tabular}
\end{center}
\end{table}

\clearpage
\begin{table}[ht]
\begin{center}
 {{\bf Table 11}. Simulation results from various estimators for the bike sharing data with $B=24$ and $N_b=17379$.}\\
\vspace{0.1in}
\small
\begin{tabular}{ccccccccccccc}
\hline
&  & \multicolumn{3}{c}{  LPRE} &  & \multicolumn{3}{c}{ CEE}\\
\cline{3-5}\cline{7-9}
&              &  est      &  sd     &  $p$-value & &  est   &  sd    &  $p$-value   \\
\hline
&  Intercept   &2.2142   &0.0280 &$<10^{-9}$          &        &2.2145   &0.0256 &$<10^{-9}$        \\
&  Workingday  &$-$0.0342&0.0102 &$8.55\times10^{-4}$ &        &$-$0.0334&0.0095 &$4.36\times10^{-4}$\\
&  Temperature &1.4525   &0.0261 &$<10^{-9}$          &        &1.4538   &0.0243 &$<10^{-9}$         \\
&  Humidity    &$-$1.1379&0.0279 &$<10^{-9}$          &        &$-$1.1404&0.0254 &$<10^{-9}$         \\
&  Windspeed   &0.1816   &0.0428 &$2.22\times10^{-5}$ &        &0.1837   &0.0405 &$5.84\times10^{-6}$   \\
\hline
&  & \multicolumn{3}{c}{  CUEE} &  & \multicolumn{3}{c}{ Renew}\\
\cline{3-5}\cline{7-9}
&              &  est      &  sd     &  $p$-value & &  est   &  sd    &  $p$-value   \\
\hline
&  Intercept   &2.2139   &0.0266 &$<10^{-9}$          &        &2.2169   &0.0263 &$<10^{-9}$        \\
&  Workingday  &$-$0.0329&0.0100 &$1.05\times10^{-3}$ &        &$-$0.0344&0.0099 &$5.01\times10^{-4}$\\
&  Temperature &1.4506   &0.0249 &$<10^{-9}$          &        &1.4507   &0.0248 &$<10^{-9}$         \\
&  Humidity    &$-$1.1377&0.0266 &$<10^{-9}$          &        &$-$1.1404&0.0263 &$<10^{-9}$         \\
&  Windspeed   &0.1812   &0.0414 &$1.22\times10^{-5}$ &        &0.1826   &0.0412 &$9.26\times10^{-6}$   \\
\hline
\end{tabular}
\end{center}
\end{table}

\clearpage
\begin{table}[ht]
\begin{center}
 {{\bf Table 12}. Simulation results from various estimators for the electric power consumption data with $B=48$ and $N_b=2049280$.}\\
\vspace{0.1in}
\small
\begin{tabular}{ccccccccccccc}
\hline
&  & \multicolumn{3}{c}{  LPRE} &  & \multicolumn{3}{c}{ CEE}\\
\cline{3-5}\cline{7-9}
&                                   &  est      &  sd     &  $p$-value & &  est   &  sd    &  $p$-value   \\
\hline
&  Intercept                        &1.1162   &0.0004 &$<10^{-9}$          &        &1.1207   &0.0004 &$<10^{-9}$        \\
&  Kitchen                          &0.2205   &0.0004 &$<10^{-9}$          &        &0.2201   &0.0004 &$<10^{-9}$        \\
&  Laundry Room                     &0.2045   &0.0005 &$<10^{-9}$          &        &0.2058   &0.0005 &$<10^{-9}$         \\
&  Water-heater and Air-conditioner &0.6326   &0.0004 &$<10^{-9}$          &        &0.6299   &0.0003 &$<10^{-9}$         \\
\hline
&  & \multicolumn{3}{c}{  CUEE} &  & \multicolumn{3}{c}{ Renew}\\
\cline{3-5}\cline{7-9}
&                                   &  est      &  sd     &  $p$-value & &  est   &  sd    &  $p$-value   \\
\hline
&  Intercept                        &1.1237   &0.0005 &$<10^{-9}$          &        &1.1160   &0.0004 &$<10^{-9}$        \\
&  Kitchen                          &0.2197   &0.0004 &$<10^{-9}$          &        &0.2205   &0.0004 &$<10^{-9}$        \\
&  Laundry Room                     &0.2041   &0.0004 &$<10^{-9}$          &        &0.2047   &0.0005 &$<10^{-9}$         \\
&  Water-heater and Air-conditioner &0.6287   &0.0004 &$<10^{-9}$          &        &0.6326   &0.0004 &$<10^{-9}$         \\
\hline
\end{tabular}
\end{center}
\end{table}

\end{document}